\documentclass[conference]{IEEEtran}
\IEEEoverridecommandlockouts
\usepackage{cite}
\usepackage{graphicx}
\usepackage{subfig}
\usepackage{multirow}
\usepackage{multicol}

\def\BibTeX{{\rm B\kern-.05em{\sc i\kern-.025em b}\kern-.08em
    T\kern-.1667em\lower.7ex\hbox{E}\kern-.125emX}}
\begin{document}

\title{A First Look at Blockchain-based Decentralized Applications
}

\author{\IEEEauthorblockN{Kaidong Wu, Yun Ma, Gang Huang, Xuanzhe Liu}
\IEEEauthorblockA{Key Lab of High-Confidence Software Technology \\
MoE (Peking University)\\
Beijing, China \\
\{wukd94, mayun, hg, liuxuanzhe\}@pku.edu.cn}
}

\maketitle

\begin{abstract}
With the increasing popularity of blockchain technologies in recent years, blockchain-based decentralized applications (DApps for short in this paper) have been rapidly developed and widely adopted in many areas, being a hot topic in both academia and industry. Despite of the importance of DApps, we still have quite little understanding of DApps along with its ecosystem. To bridge the knowledge gap, this paper presents the first comprehensive empirical study of blockchain-based DApps to date, based on an extensive dataset of 995 Ethereum DApps and 29,846,075 transaction logs over them. We make a descriptive analysis of the popularity of DApps, summarize the patterns of how DApps use smart contracts to access the underlying blockchain, and explore the worth-addressing issues of deploying and operating DApps. Based on the findings, we propose some implications for DApp users to select proper DApps, for DApp developers to improve the efficiency of DApps, and for blockchain vendors to enhance the support of DApps.
\end{abstract}


\section{Introduction}

Since the invention of the cryptocurrency Bitcoin by Satoshi Nakamoto in 2008 \cite{nakamoto2008bitcoin}, blockchain technologies have been rapidly developed, and have drawn lots of attentions from both academia and industry. A blockchain is a decentralized, distributed, and public digital ledger that is used to record transactions across many nodes so that any involved record cannot be altered retroactively, without the alteration of all subsequent blocks. Due to its advantages of decentralization, immutability, security, and transparency, the blockchain has become one of the most promising infrastructural technologies for the next generation of Internet-based systems, such as public services, Internet of Things (IoT), reputation systems, and security services \cite{zheng2018blockchain}.

Essentially, the blockchain is a kind of distributed systems, providing the computation capability for applications to run on multiple computation nodes. Since blockchains have no centralized control but are maintained according to decentralized consensus models, applications on blockchains actually belong to decentralized applications (DApp for short), a special type of software where the application execution is not controlled by a single entity. In history, DApps usually refer to applications that run on the Peer-to-Peer (P2P) network of computers rather than a single centralized computer. Many famous DApps have ever been developed and widely spread, such as BitTorrent \cite{bittorrent} for file sharing, BitMessage \cite{bitmessage} for instant messaging, and Popcorn Time \cite{popcorntime} for video streaming.

Blockchains provide a general computation abstraction via the mechanism of smart contracts, making it easy to develop DApps for various application contexts. For example, the Ethereum blockchain \cite{ethereumproject} provides Turing-complete smart contracts for developers to implement general-purpose programs. In consequence, with the prosperity of blockchains, more and more blockchain-based DApps have emerged, being adopted to almost all areas. According to a recent report, the value of the biggest blockchain-based DApp market, Ethereum DApp market, has reached billions of dollars as of Jan. 2019 \cite{dappreport}.

Despite of the popularity of blockchain-based DApps, there is no comprehensive understanding of such a rising ecosystem. The industrial reports on blockchain-based DApps focus on only basic usage statistics \cite{mediumreport, hackernoonreport}, such as the number of daily active users and the amount of transactions. Academic research efforts are mainly devoted on the underlying blockchain system \cite{gervais2016security, zheng2018detailed, wang2019monoxide} as well as the mechanism of smart contracts \cite{luu2016making, kosba2016hawk}. Few studies investigate the characteristics and development practices of blockchain-based DApps.

To bridge the knowledge gap, in this paper, we conduct the first comprehensive empirical study on blockchain-based DApps. We choose Ethereum, the largest and the most popular platform for running blockchain-based DApps, as our target. We select 995 popular Ethereum DApps, which contain 5,158 smart contracts, and retrieve all the corresponding 29,846,075 transactions occurred in 2018. Based on the dataset, we try to answer the following three research questions:\\

\begin{itemize}
\item \textbf{RQ1: How is the popularity of DApps distributed?}
We explore the popularity of DApps by the number of unique users, transactions, and transaction volumes, and compare categories of DApps. We also examine the change of popularity as time evolves. By answering this question, we can provide an overview of the DApp market for all stakeholders in the DApp ecosystem.\\
\item \textbf{RQ2: Are there any common practices of developing DApps?}
We investigate whether DApps are open source and how smart contracts are organized in a DApp. By answering this question, we can reveal the development practices of current DApps. \\ 
\item \textbf{RQ3: How much is the cost of DApps when running on the blockchain?}
Running DApps have to pay fee to blockchain miners for deploying and executing smart contracts. By answering this question, we can provide some recommendations for developers to reduce the cost of DApps.
\end{itemize}

Our key findings are summarized as follows:

\begin{itemize}
\item \textbf{Popularity of DApps} (Sections \ref{sec:distribution} \& \ref{sec:categories}).
The distributions of the number of users, transactions, and transaction volumes against the popularity of DApps typically follow the Pareto principle, i.e., a few DApps have substantial popularity. DApps with financial features have large influence on the market.\\

\item  \textbf{Growth of DApps} (Section \ref{sec:growth}).
The number of DApps was first rapidly grown for the categories including Exchanges, Gambling, and Finance. As more DApps are developed, the number of DApps from the high-risk category increases significantly, leading to potential security issues.\\

\item \textbf{Open source of DApps} (Section \ref{sec:open_source}).
Currently, the open-source levels of DApps are a bit diverse and not satisfactory. Only 15.7\% of DApps are fully open source where the code of both the DApp and related smart contracts are available. In contrast, 25.0\% of DApps are completely closed source. In general, DApps whose smart contracts are open source, usually have more transactions than others, indicating that open source can have potentially significant impacts on the popularity of DApps.\\

\item \textbf{Usage patterns of smart contracts} (Section \ref{sec:patterns}).
75\% of DApps consist of only one smart contract. For DApps with multiple smart contracts, there are 3 usage patterns of smart contracts: \textit{leader-member} where smart contracts are invoked with each other through internal transactions, \textit{equivalent} where there is no invocation between smart contracts, and \textit{factory} where the child contracts are deployed by a factory contract.\\

\item \textbf{Cost of deploying smart contracts} (Section \ref{sec:deploy_cost}).
The average deployment cost per smart contract for single-contract DApps is less than that for multi-contract DApps. The deployment cost per smart contract is correlated with the line of code (LoC) and the number of functions (NoF) where NoF has more influence on the deployment cost.\\

\item \textbf{Cost of executing smart contracts} (Section \ref{sec:exec_cost}).
In the median case, only 50\% of the prepaid fee for executing smart contracts is actually used, leading to half of the prepaid fee being locked in transactions until they are confirmed. Contract executions with internal transactions cost more than those without internal transactions.

\end{itemize}

To the best of our knowledge, our work is the first comprehensive analysis of blockchain-based DApps to date. Our findings present an overview of Ethereum DApp market, motivating future research and development. Specifically, the findings show strong implications for multiple stakeholders of the market. For example, end-users can choose suitable DApps according to DApp distribution; end-users and developers can set proper amount of prepaid fee for executing smart contracts to avoid their assets being locked during contract executions; DApp developers can understand DApp life cycle better and use the suitable pattern to design the DApp architecture and smart contracts; blockchain vendors can optimize some mechanisms to serve DApps better.

The remaining of this paper is organized as follows. We introduce the background of the Ethereum blockchain and Ethereum DApps in Section \ref{sec:background}. We describe our dataset in Section \ref{sec:dataset}. We characterize the DApp popularity in Section \ref{sec:descriptive}, analyze features in the development of DApps in Section \ref{sec:develop}, and investigate the cost of deploying and executing DApps in Section \ref{sec:sc}. We summarize our findings and implications in Section \ref{sec:implications}. We survey related work in Section \ref{sec:related_work}, and conclude the paper in Section \ref{sec:conclusion}.

\section{Background}
\label{sec:background}

We choose the largest DApp market, Ethereum DApps, to conduct our empirical study. In this section, we give some background knowledge of the Ethereum blockchain and the Ethereum DApps.

\subsection{Ethereum Blockchain}

The blockchain is a ledger system based on a P2P network, keeping records of transactions which represent value transfers between accounts. In the network, all nodes receive transactions, pack them into a block that is linked to the previous block, and broadcast the block. According to the consensus mechanism, if most nodes receive and accept a block, then the block will be a part of the ledger.

Ethereum is the second-generation blockchain after Bitcoin.
The ledger of Ethereum is used to support the cryptocurrency, Ether (ETH).

In Ethereum, there are two kinds of accounts: user accounts and smart-contract accounts.
The user accounts represent participants, including callers (who call functions of smart contracts), deployers (who deploy smart contracts on Ethereum), and miners (whose nodes work to do contribution to the ledger).
The smart-contract accounts represent the smart contract which is a type of programs that are saved in and able to run on blockchains, called chaincode (code on chain) as well \cite{szabo1994smart, szabo1996smart}. Ethereum is the first blockchain providing Turing-complete programming language to develop smart contracts.

Transactions can be classified from two dimensions:

On the one hand, according to the \textit{data in the transaction}, transactions can be divided into Ether transfers and contract executions. An Ether transfer represents that a user account transfers some Ether to another one. A contract execution represents that an account calls a function of a smart contract with some data as the input and some Ether as the fee for executing the contract.

On the other hand, according to the \textit{transaction initiator}, transactions can be divided into external transactions which are initiated by user accounts, and internal transactions which are initiated by smart contract accounts.

\begin{figure}[htbp]
    \centerline{
    \includegraphics[width=0.95\linewidth]{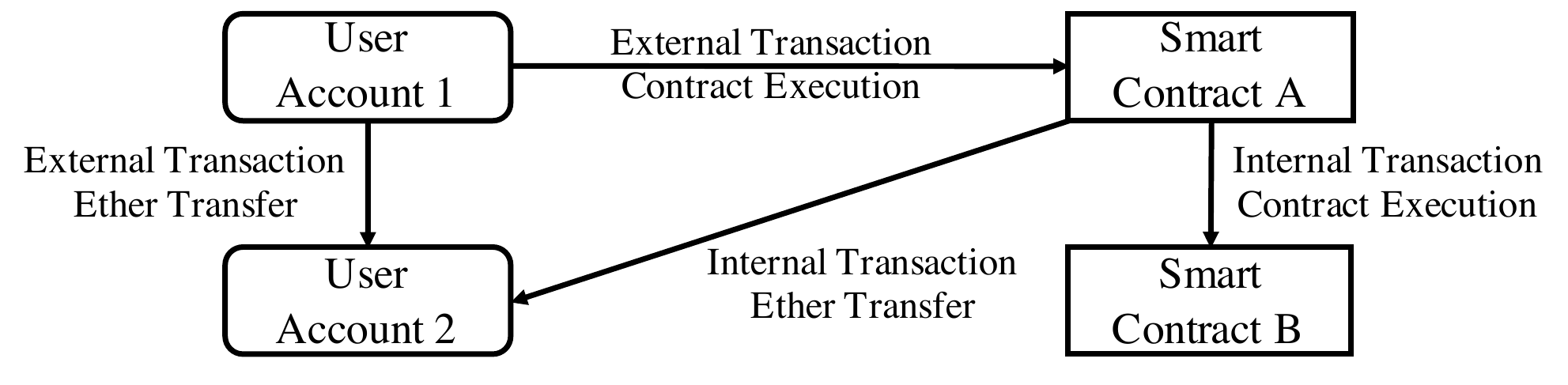}}
    \caption{Relationship between the two classifications of transactions.}
    \label{fig:kind-trans}
\end{figure}

Figure \ref{fig:kind-trans} shows the relationship between the two classifications. If the target account of a transaction is a user account, the transaction belongs to Ether transfer. If the target account of a transaction is a smart-contract account, the transaction belongs to contract execution. 

Accounts have to pay fee for all transactions. These fees are uniformly called gas in the Ethereum ecosystem.

\subsection{Ethereum DApp}

\begin{figure}[htbp]
	\centerline{
    \subfloat[Direct.]{
        \includegraphics[width=0.45\linewidth]{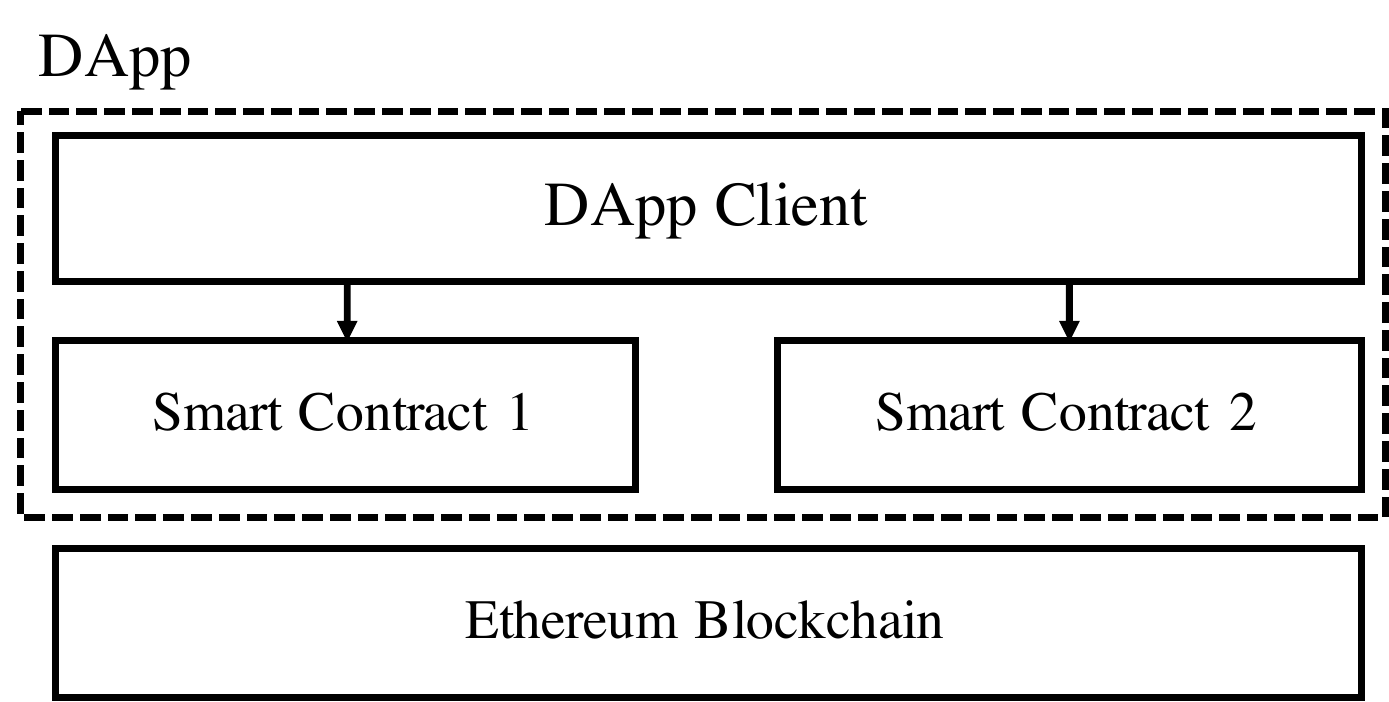}
        \label{fig:archit-direct}
    }
    \subfloat[Indirect.]{
        \includegraphics[width=0.45\linewidth]{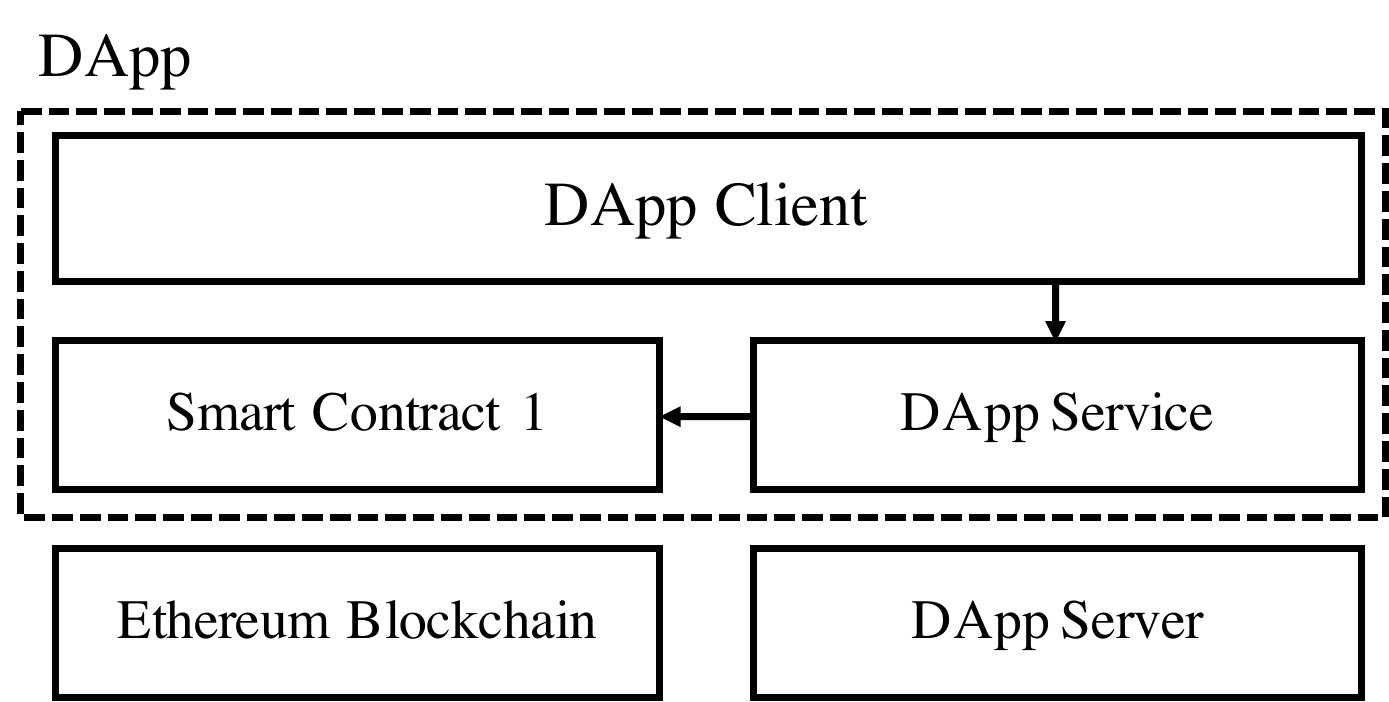}
        \label{fig:archit-indirect}
    }}
    \centerline{
    \subfloat[Mixed.]{
        \includegraphics[width=0.45\linewidth]{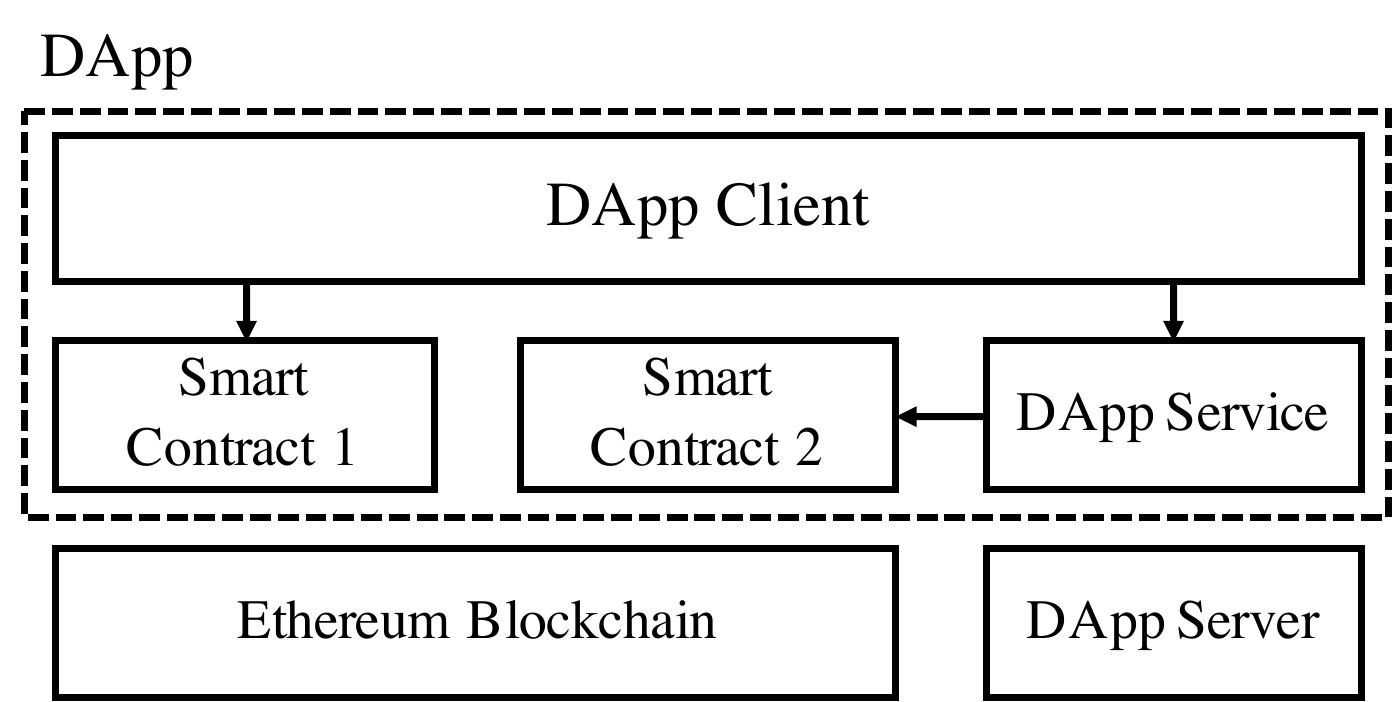}
        \label{fig:mixed}
    }}
	\caption{Three kinds of DApp architectures.}
	\label{fig:archit}
\end{figure}

Ethereum blockchain provides computation and storage capabilities via the mechanism of smart contracts. Therefore, Ethereum DApps can deploy smart contracts to use the capabilities provided by Ethereum to implement business logics. In theory, all the processes and data of a blockchain-based DApp should be handled and stored on the blockchain for pure decentralization. However, due to the performance bottleneck of state-of-the-art blockchain systems, current DApps usually implement only parts of their functionality on the blockchain. As a result, three kinds of architectures are adopted by Ethereum DApps in practice as shown in Figure \ref{fig:archit}: direct, indirect and mixed. For DApps of the direct architecture (Figure \ref{fig:archit-direct}), the client directly interacts with smart contracts deployed on the Ethereum. DApps of the indirect architecture (Figure \ref{fig:archit-indirect}) have back-end services running on a centralized server, and the client interacts with smart contracts through the server. DApps of the mixed architecture combines the preceding two architectures where the client interacts with smart contracts both directly and indirectly through a back-end server.

Solidity is the programming language for developing smart contracts in the Ethereum community. It is a JavaScript-like language, in which there are contracts (like classes), functions and events. The source code of a smart contract is compiled into bytecode to be deployed on Ethereum. After deployment, the smart contract will get an address.

\textbf{Smart contract deployment.}
All accounts can deploy smart contracts. Sometimes developers use a smart contract to deploy other child contracts. The usage pattern is called "factory pattern", and will be discussed in Section \ref{sec:patterns}.

\textbf{Cost of smart contracts.} As mentioned above, accounts have to pay gas for every transaction. Cost of a smart contract consists two parts: deployment and contract execution. The deployment of a smart contract can be seen as a contract execution of calling a special function \textit{constructor()}. In this paper, we separate the deployment cost out of the general execution cost to study the two kinds of costs respectively. The deployment cost can represent the complexity of the contract. Since executions are uniformly encoded in transactions, and finally packed into blocks, total gas sent in transactions of a block is limited. So, a contract execution that costs more gas than the limit will fail and change nothing.
\section{The dataset}
\label{sec:dataset}

In this section, we introduce the dataset that we collect to conduct our empirical study.

Since there is no official app marketplace for blockchain-based DApps like App Store for iOS apps or Google Play for Android apps, we resort to third-party collections of DApps. Specifically, we choose DApps from \textit{State of the DApps} \cite{stateofthedapps}, which is a privately funded and independent website that collects DApps from major blockchain systems. Established in 2017, \textit{State of the DApps} has grown to be the main directory of DApps where DApp developers can submit their DApps to get published. It is also worth to mention that \textit{State of the DApps} is referenced by the official homepage of Ethereum. So Ethereum DApps from this site are representative to understand the whole ecosystem.

We retrieve all the 1,749 Ethereum DApps that can be found on \textit{State of the DApps} as of Jan. 2019. For each DApp, we collect the following information.\\

\begin{itemize}
    \item \textbf{Basic information of the DApp}, including the name, the category it belongs to, and the first publishing time.\\
    
    \item \textbf{Smart contracts which the DApp consists of.} \textit{State of the DApps} allows developers to specify the addresses of smart contracts that the published DApp consists of. Therefore, we can retrieve such information for some DApps. Then, we try to obtain the source code of each smart contract on \textit{Etherscan} \cite{etherscan}, which is a block explorer and analytic platform of Ethereum where developers can submit the source code of their smart contracts. \\
    
    \item \textbf{Transactions related to the DApp.}
    We get blocks submitted in 2018 from Ethereum blockchain, and retrieve related contract executions in these blocks by checking addresses of participants. For each transaction, we extract addresses of senders and receivers, gas sent, input data, status (whether the contract execution succeeds) and how many Ethers transferred. Related internal transactions and gas used of external transactions are extracted by simulated executions, and double checked by being compared with results from Etherscan.
\end{itemize}

\begin{table}[htbp]
    \caption{Statistics of the dataset used in the empirical study.}
    \begin{center}
        \begin{tabular}{lr}
            \hline
            \textbf{\textit{DApps}} & 995 \\
            \textbf{\textit{Smart contracts}} & 5,158 \\
            \textbf{\textit{Open-source smart contracts}} & 2,568 \\
            \textbf{\textit{Transactions}} & 29,846,075 \\
            \textbf{\textit{User Accounts}} & 2,199,059 \\
            \textbf{\textit{Transaction volume (ETH)}} & 9,057,344.360 \\
            \hline
        \end{tabular}
        \label{tab:dataset}
    \end{center}
\end{table}

In order to deeply investigate how DApps utilize the capabilities provided by the blockchain, we filter those DApps that developers do not provide the addresses of smart contracts for.
Because we cannot retrieve smart contracts just by DApps' information and analyze how they work.
After filtering, our dataset has 995 DApps left, which are used for the study in the following sections. Table \ref{tab:dataset} shows the statistics of our dataset.

We get 5,158 smart contracts from lists that DApps provide, among which 2,568 contracts' source code can be retrieved. There are 29,846,075 related transactions initiated in 2018, by which 2,199,059 user accounts transfer 9,057,344,360 Ethers.
\section{Popularity of DApps}
\label{sec:descriptive}

In this section, we try to describe an overview of the Ethereum DApp market, and answer RQ1, i.e., how is the popularity of DApps distributed? We first study the popularity by some metrics, and then compare DApps in different categories. Finally, we examine the growth of the number of DApps as time evolves.

\subsection{DApp Distribution}
\label{sec:distribution}

We use three metrics to measure the popularity of DApps: (1) the number of unique addresses (user accounts); (2) the number of transactions; (3) the amount of transaction volume (in ETH).

\subsubsection{Popularity by Users}

\begin{figure}[htbp]
	\centerline{
    \subfloat[Percentage of users against DApp rank.]{
        \includegraphics[width=.45\linewidth]{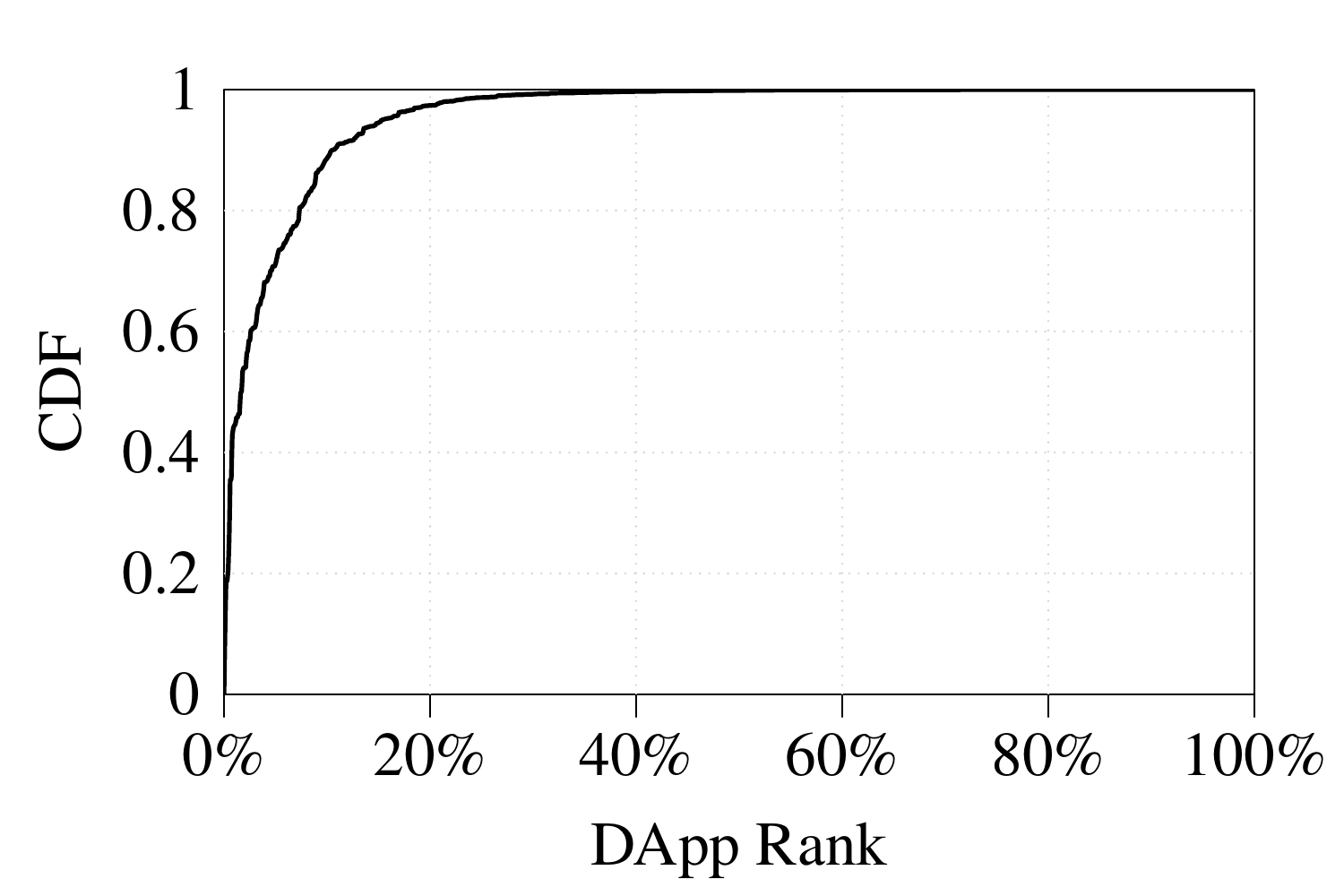}
        \label{fig:rank_users}
    }
    \subfloat[Users of a DApp.]{
        \includegraphics[width=0.45\linewidth]{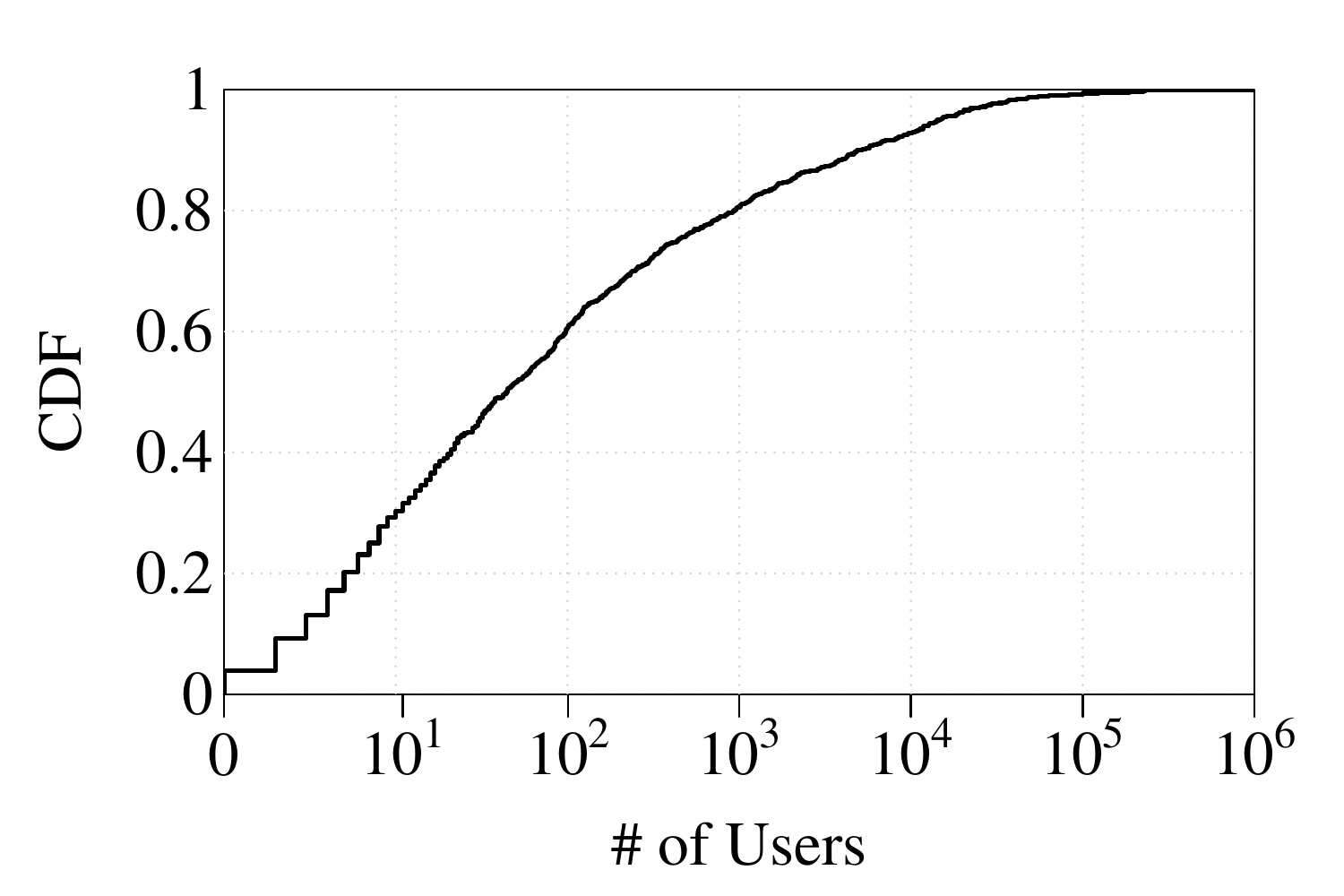}
        \label{fig:cdf_users}
    }}
	\caption{DApp popularity by users.}
\end{figure}

We extract all the user accounts occurred in the transactions of a DApp as the number of users that the DApp has.

Figure \ref{fig:rank_users} shows the cumulative distribution function (CDF) of the percentage of DApp users against the DApp ranking by transactions. We can see that the DApp users follow the Pareto principle, i.e. less than 20\% of DApps have almost all users. We can conclude that the more a DApp is used, the more users it has.

We also explore the distribution of users of each DApp. Figure \ref{fig:cdf_users} indicates that about 80\% of DApps are used by only less than 1000 users.

\subsubsection{Popularity by Transactions}

\begin{figure}[htbp]
	\centerline{
    \subfloat[Percentage of transactions against DApp rank.]{
        \includegraphics[width=.45\linewidth]{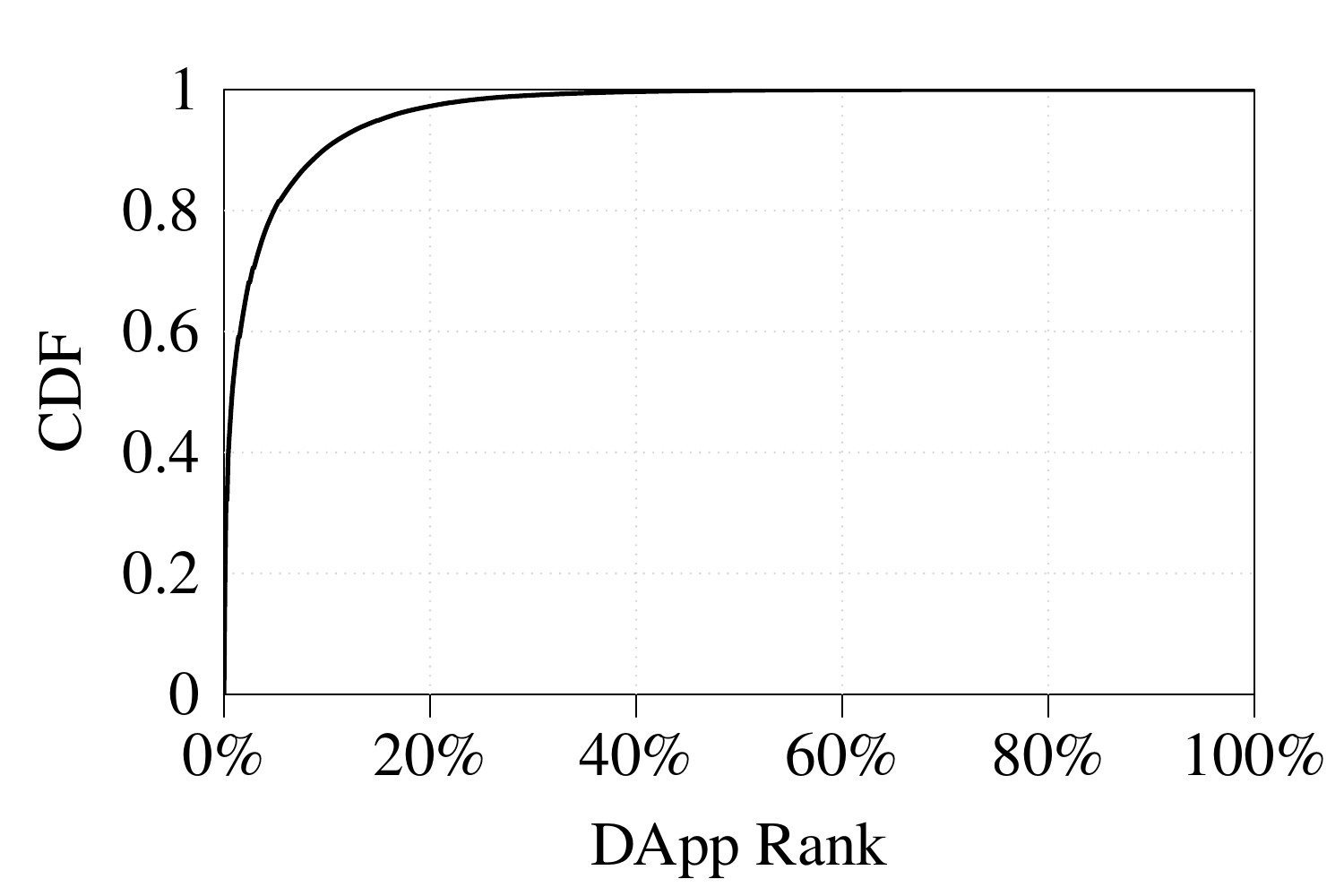}
        \label{fig:rank_trans}
    }
    \hfill
    \subfloat[Transactions of a DApp.]{
        \includegraphics[width=.45\linewidth]{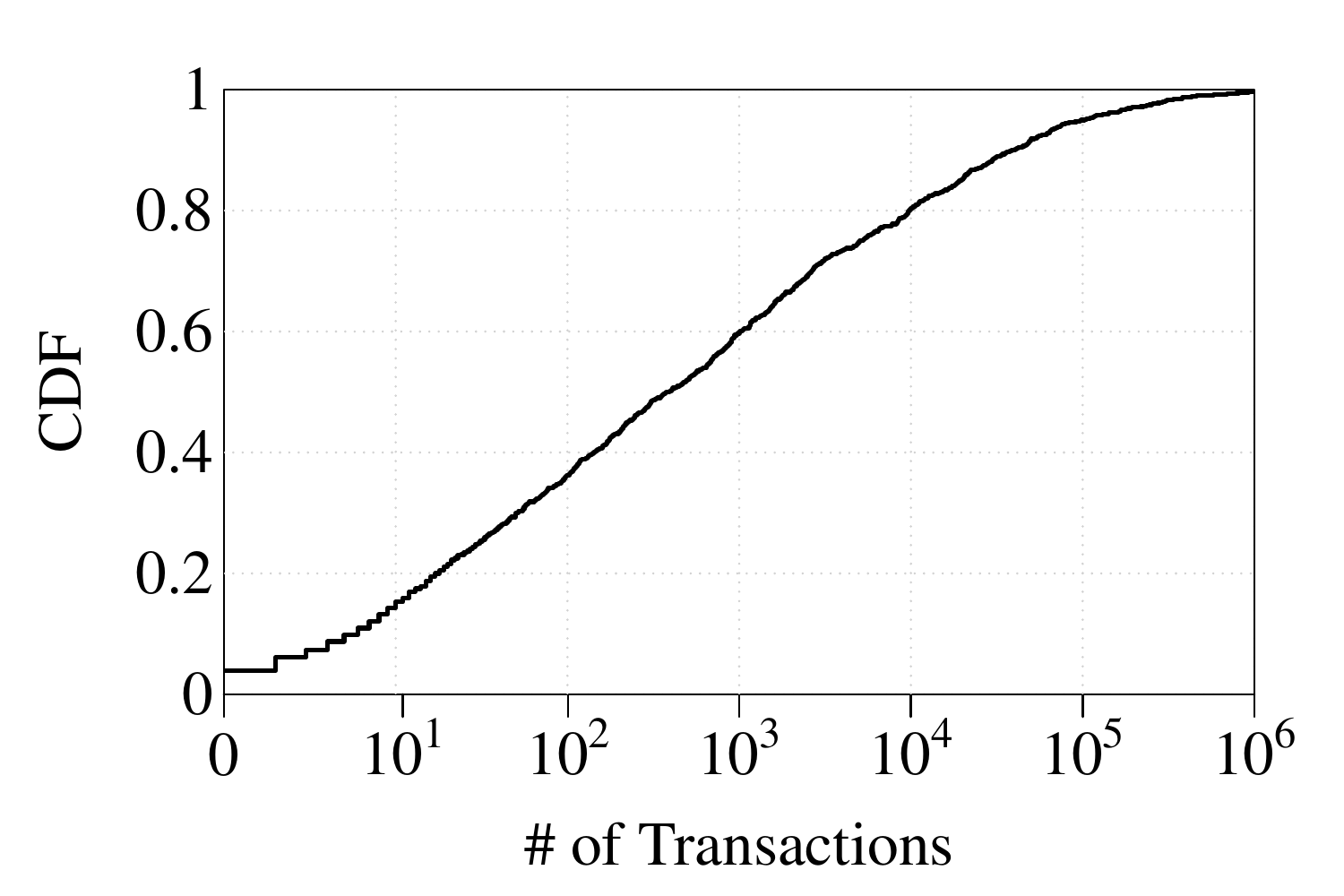}
        \label{fig:cdf_trans}
    }}
    \caption{DApp popularity by transactions.}
\end{figure}

Because all transactions are added to the block with paying gas, transactions can be seen as real behaviours of users. Figure \ref{fig:rank_trans} shows that the CDF of the percentage of transactions of DApps against DApp ranking by transactions. We can find that a few (about 5\%) DApps have 80\% of transactions. It follows the Pareto principle as well.

Figure \ref{fig:cdf_trans} shows that over 80\% of DApps have less than 10,000 transactions. Such findings indicate a "long-tail" of DApps that are rarely active.\\

\subsubsection{Popularity by Transaction Volumes}

\begin{figure}[htbp]
	\centerline{
    \subfloat[Percentage of transaction volumes against DApp rank.]{
        \includegraphics[width=.45\linewidth]{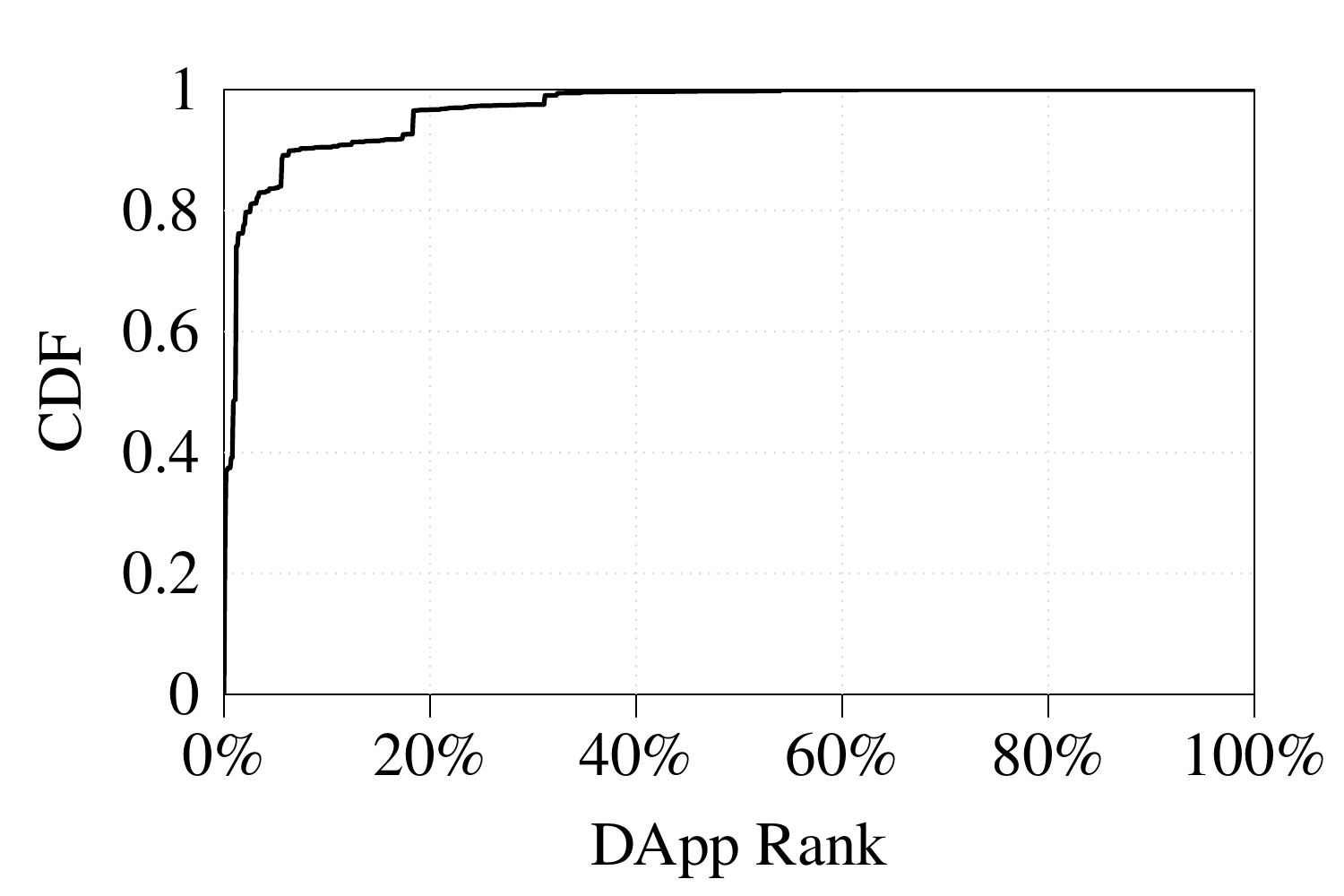}
        \label{fig:rank_volume}
    }
    \hfill
    \subfloat[Transaction volume of a DApp.]{
        \includegraphics[width=.45\linewidth]{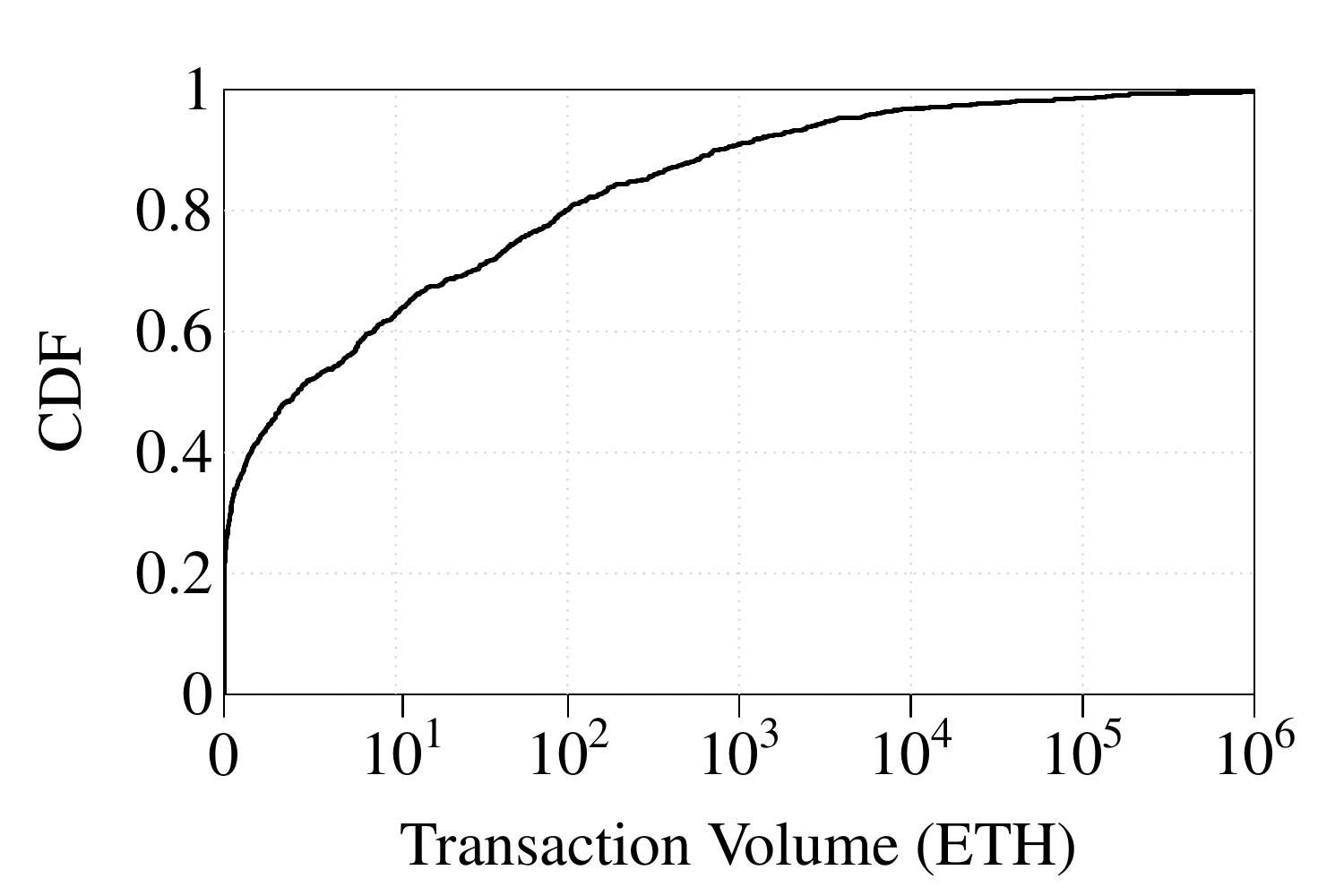}
        \label{fig:cdf_volume}
    }}
    \caption{DApp popularity by ETH trade.}
\end{figure}

In Ethereum, each transaction needs to pay gas, charged by Ethers, to miners, but they can transfer 0 Ether. Therefore, DApps having many transactions may not be economic beneficial. So we study the actual transaction volume of DApps. Figure \ref{fig:rank_volume} illustrate the CDF of the percentage of DApps' transaction volumes against the DApp rank. We can find that the Pareto principle also holds for DApp transaction volumes.

Figure \ref{fig:cdf_volume} shows that about 60\% of DApps have only less than 10 Ethers (about 6,000 dollars in 2018) a year. 27.25\% of DApps get few Ethers.

\begin{table*}[htbp]
\caption{DApp popularity by categories.\\ (Categories are sorted by Transactions)}
\begin{center}
    \begin{tabular}{lrrrrrrrr}
        \hline
        \multicolumn{1}{c}{\multirow{2}{*}{\textbf{Category}}} & \multicolumn{2}{c}{\multirow{1}{*}{\textbf{DApps}}} & \multicolumn{2}{c}{\multirow{1}{*}{\textbf{Users}}} & \multicolumn{2}{c}{\multirow{1}{*}{\textbf{Transactions}}} & \multicolumn{2}{c}{\multirow{1}{*}{\textbf{Transaction Volume}}} \\
        & \multicolumn{1}{c}{\textbf{\#}} & \multicolumn{1}{c}{\textbf{\%}} & \multicolumn{1}{c}{\textbf{\#}} & \multicolumn{1}{c}{\textbf{\%}} & \multicolumn{1}{c}{\textbf{\#}} & \multicolumn{1}{c}{\textbf{\%}} & \multicolumn{1}{c}{\textbf{\#}} & \multicolumn{1}{c}{\textbf{\%}} \\
        \hline
        \textbf{\textit{Exchanges}} & 71 & 7.1\% & 778,031 & 35.4\% & 13,708,713 & 45.9\% & 5,570,026.10 & 61.5\% \\
        \textbf{\textit{Games}} & 294 & 29.5\% & 184,730 & 8.4\% & 5,834,574 & 19.5\% & 226,506.11 & 2.5\% \\
        \textbf{\textit{Finance}} & 93 & 9.3\% & 517,525 & 23.5\% & 2,571,729 & 8.6\% & 2,321,199.88 & 25.6\% \\
        \textbf{\textit{Gambling}} & 154 & 15.5\% & 77,790 & 3.5\% & 1,781,856 & 6.0\% & 448,476.51 & 5.0\% \\
        \textbf{\textit{Development}} & 30 & 3.0\% & 269,821 & 12.3\% & 1,154,346 & 3.9\% & 20,525.36 & 0.2\% \\
        \textbf{\textit{Storage}} & 13 & 1.3\% & 249,544 & 11.3\% & 1,031,779 & 3.5\% & 11.29 & 0.0\% \\
        \textbf{\textit{High-risk}} & 130 & 13.1\% & 47,186 & 2.1\% & 965,131 & 3.2\% & 370,543.94 & 4.1\% \\
        \textbf{\textit{Wallet}} & 17 & 1.7\% & 193,790 & 8.8\% & 787,160 & 2.6\% & 2,020.50 & 0.0\% \\
        \textbf{\textit{Governance}} & 18 & 1.8\% & 188,201 & 8.6\% & 633,211 & 2.1\% & 132.29 & 0.0\% \\
        \textbf{\textit{Property}} & 24 & 2.4\% & 46,707 & 2.1\% & 485,341 & 1.6\% & 47,421.00 & 0.5\% \\
        \textbf{\textit{Identity}} & 12 & 1.2\% & 82,251 & 3.7\% & 425,860 & 1.4\% & 5,330.45 & 0.1\% \\
        \textbf{\textit{Media}} & 48 & 4.8\% & 127,315 & 5.8\% & 403,055 & 1.4\% & 1,144.43 & 0.0\% \\
        \textbf{\textit{Social}} & 72 & 7.2\% & 88,355 & 4.0\% & 381,534 & 1.3\% & 10,065.19 & 0.1\% \\
        \textbf{\textit{Security}} & 14 & 1.4\% & 40,684 & 1.9\% & 127,550 & 0.4\% & 17,211.19 & 0.2\% \\
        \textbf{\textit{Energy}} & 3 & 0.3\% & 21,312 & 1.0\% & 95,025 & 0.3\% & 22,127.17 & 0.2\% \\
        \textbf{\textit{Insurance}} & 1 & 0.1\% & 5,755 & 0.3\% & 19,575 & 0.1\% & 0.52 & 0.0\% \\
        \textbf{\textit{Health}} & 2 & 0.2\% & 4 & 0.0\% & 9 & 0.0\% & 0.00 & 0.0\% \\
        \textbf{\textit{All DApps}} & 995 &  & 2,199,059 &  & 29,846,075 &  & 9,057,344.36 &  \\
        \hline
    \end{tabular}
    \label{tab:category_distribution}
\end{center}
\end{table*}

\subsection{DApp Category}
\label{sec:categories}

In \textit{State of the DApps}, DApps are divided into 17 categories: Exchanges, Games, Finance, Gambling, Development, Storage, High-risk, Wallet, Governance, Property, Identity, Media, Social, Security, Energy, Insurance and Health.
The categorization is similar to that of mobile apps in Google Play \cite{google_play}, but simpler and rougher.

Some categories are not used to categorize mobile apps, such as Finance and Exchanges, which are two typical categories of DApps.
Because of the financial feature of Ethereum, cryptocurrencies (category Finance), cryptocurrency management services (category Wallet) and cryptocurrency markets (category Exchanges) are hot. Recently, smart contracts are used in gambling (category Gambling), collectible card games and gambling games (category Games).

DApps from category Property are marketplaces and services for other products, like software. In other categories, DApps claim they provide distributed or decentralized services to meet users' requirements.

DApps from category Development claim that they provide services to help to use distributed computing services of blockchain and develop smart contracts and DApps. Category Storage is used to mark DApps that provide decentralized storage services, Compared with similar mobile apps, decentralization and robustness is placed on them.

High-risk seems to be a tag rather than a category. According to \cite{chen2018detecting}, there are over 400 Ponzi schemes running on Ethereum. Therefore, category High-risk marks DApps in which users may take high-risk, like investment traps and Ponzi Scheme games. 

Table \ref{tab:category_distribution} shows the popularity of DApps among different categories. 
We can find that category Games has the most DApps (29.5\%) but 8.4\% of users (7th), the second most (19.5\%) transactions and 2.5\% of transaction volume (5th). DApps of this category have become a hot topic in Ethereum DApp market.
Category Exchanges and Finance have many users (35.4\% and 23.5\%) and high ratio of transaction volume (61.5\% and 25.6\%). These DApps have great influence.
Category Identity, Media, Social, Security, Energy, Insurance and Health have fewer users and transactions. They represent new designs of DApps, but their practicability still need verification.

\subsection{Growth over Time}
\label{sec:growth}

\begin{figure}[htbp]
	\centerline{
    \subfloat[Growth of all DApps.]{
        \includegraphics[width=0.95\linewidth]{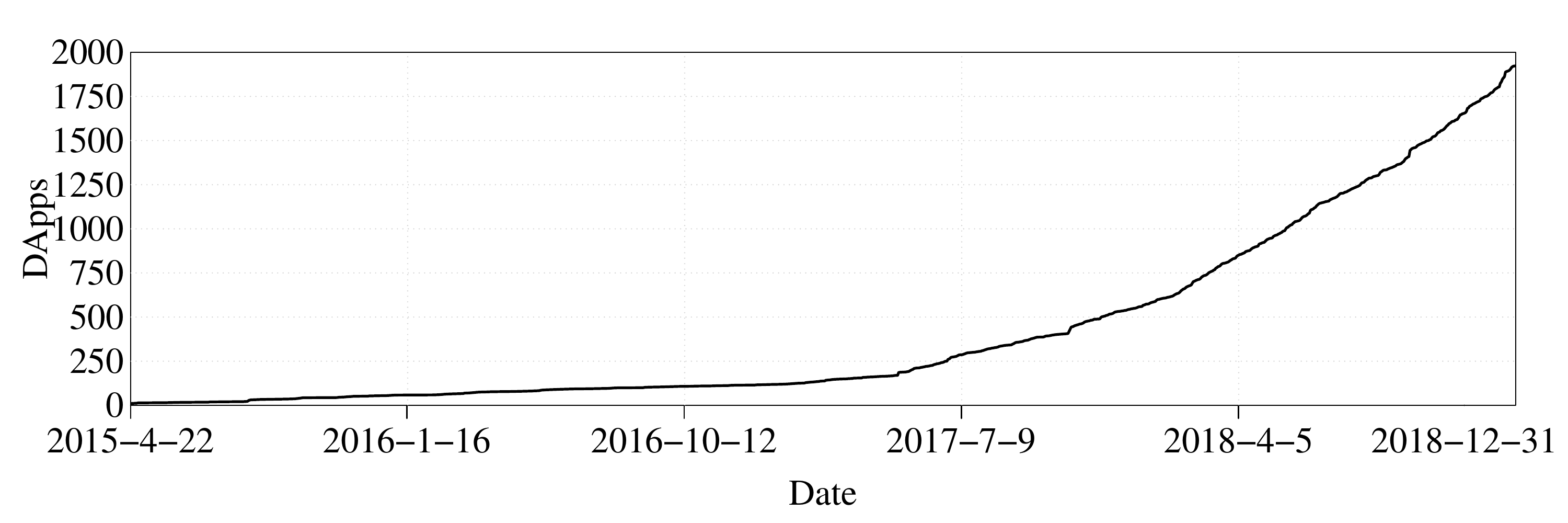}
        \label{fig:all_dapp_grow}
    }
    }
    \centerline{
    \subfloat[Growth of DApps from category Exchanges, Games, Finance, Gambling, Development and High-risk.]{
        \includegraphics[width=0.95\linewidth]{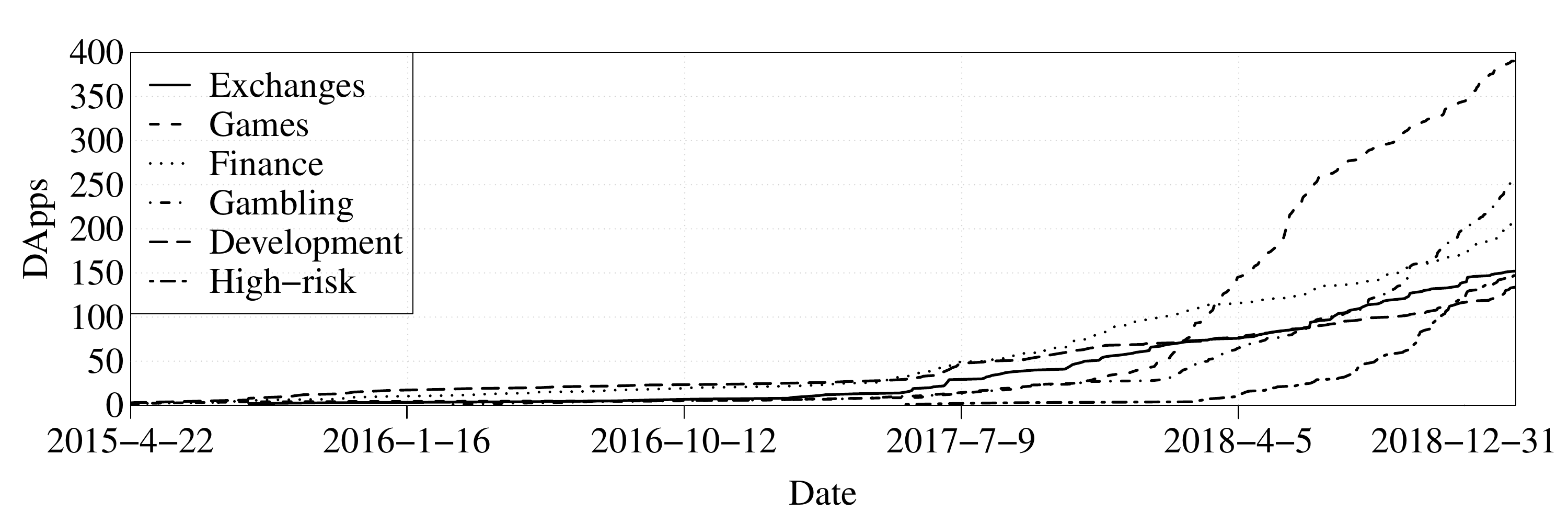}
        \label{fig:category_grows}
    }}
	\caption{DApps' growth over time.}
\end{figure}

Figure \ref{fig:all_dapp_grow} shows how the number of DApps grew over time before Jan. 1st, 2019. In our dataset, the first DApp was published at April 22, 2015. The number of DApps grew significantly from May 2017. After Feb. 2018, more DApps were published every day.

Figure \ref{fig:category_grows} shows the growth of DApps of top six categories ranked by transactions. 
DApps from category Development and Finance were first published.
At the beginning, all categories had similar trends. After May 16, 2017, the numbers of DApps from category Exchanges, Games, Gambling and Finance grew rapidly. In Feb. 2018, category Games had similar number with category Exchanges and Finance, then grew over the other two and led to the significant growth of the number of all DApps.

The first DApp from category High-risk appeared in May 2017. In Feb. 2018, the number of DApps of High-risk grew more rapidly, of which the trend was similar with category Gambling's. At Dec 31st, 2018, there were 154 Gambling DApps and 130 High-risk DApps in Ethereum DApp market.

So, we can conclude that DApps from category Exchanges and Finance were hot early, and then DApps with the financial feature like gambling games grew rapidly. But at the same time, high-risk DApps also became more and more.
\section{Development of DApps}
\label{sec:develop}

In this section, we dig deeply into DApps and the corresponding smart contracts to answer RQ2, i.e., are there any common practices of developing DApps? We first study whether DApps are open source, being able to be audited by any blockchain participant. Then we investigate how developers implement smart contracts to support DApps.

\subsection{Open Source}
\label{sec:open_source}

As illustrated in Figure \ref{fig:archit}, a DApp can be divided into two parts: on-chain part where smart contracts are implemented to use the capability of blockchains, and off-chain part where traditional programs are implemented to provide services to end-users. Thus, we study two levels of open source of DApps: contract level (on-chain) and project level (off-chain).

Although all smart contracts can be directly retrieved from the blockchain, they are only in the form of bytecode which is not readable. In practice, developers can submit the source code of smart contracts to block explorers or open-source code repositories. For the other parts of DApps, they can be open source through classical open-source channels, like GitHub.

According to whether DApps are open source at the contract level and the project level, there are 6 categories of the open-source degree of DApps, as shown in Table \ref{tab:open_source_categories}. Note that smart contracts may be partially open source, meaning that developers submit the source code of only part of smart contracts consisting of a DApp.

\begin{table}[htbp]
\newcommand{\tabincell}[2]{\begin{tabular}{@{}#1@{}}#2\end{tabular}}
\caption{Categories of the open-source degree of DApps.}
\begin{center}
\begin{tabular}{lcc}
\hline
& \textbf{\tabincell{l}{The DApp project\\is open source}} & \textbf{\tabincell{l}{The DApp project\\isnot open source}} \\
\hline
\textbf{\textit{\tabincell{l}{Smart contracts\\are all open source}}} & GA & NA \\
\hline
\textbf{\textit{\tabincell{l}{A part of smart contracts\\are open source}}} & GP & NP \\
\hline
\textbf{\textit{\tabincell{l}{Smart contract\\are all closed source}}} & GC & NC \\
\hline
\end{tabular}
\label{tab:open_source_categories}
\end{center}
\end{table}

\begin{figure}[htbp]
    \begin{minipage}[htbp]{.45\linewidth}
    \centerline{
    \includegraphics[width=.95\linewidth]{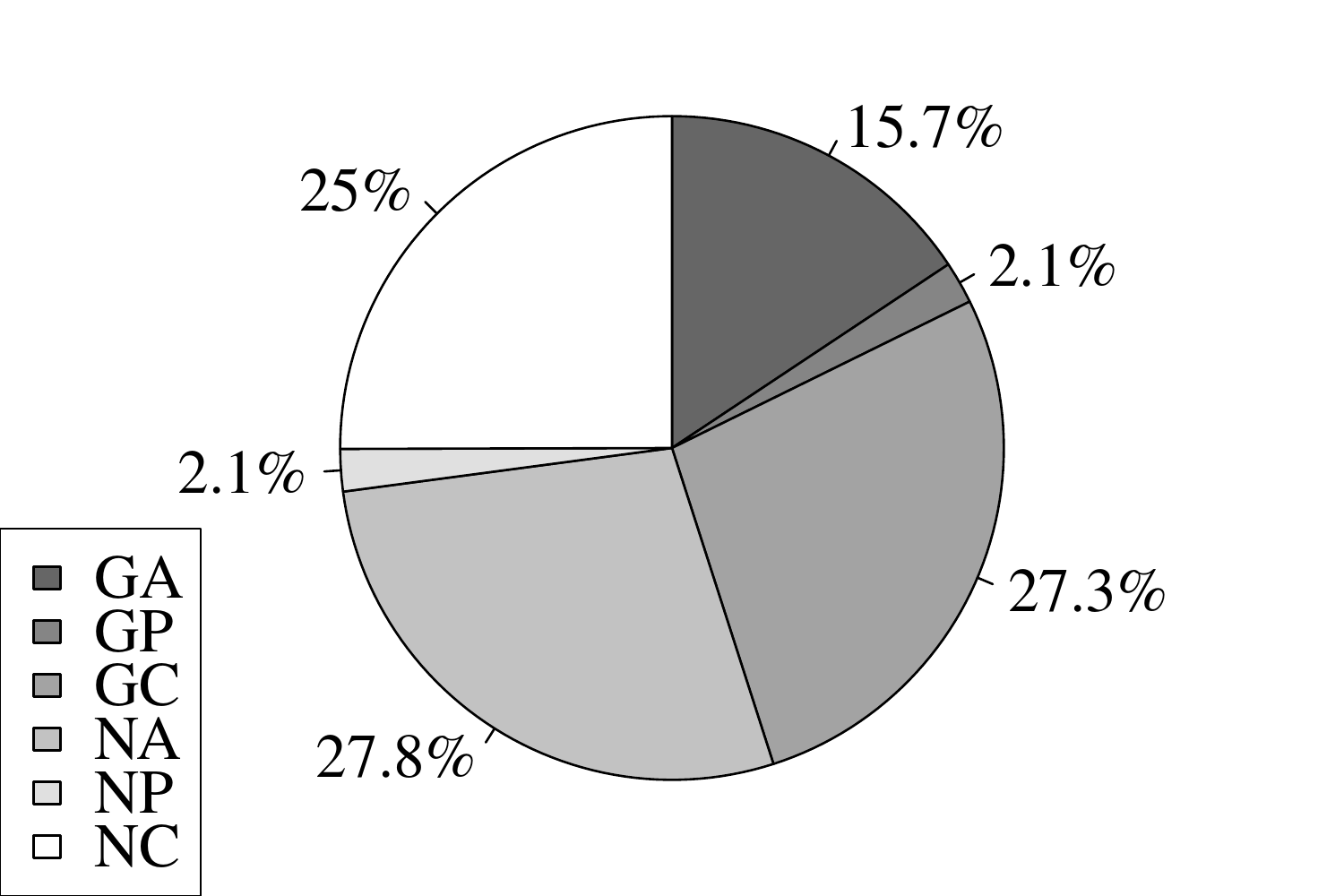}
    }
    \caption{Percentage of open-source DApps.}
    \label{fig:open_source}
    \end{minipage}
    \hspace{0.05\linewidth}
    \begin{minipage}[htbp]{.45\linewidth}
    \centerline{
    \includegraphics[width=.95\linewidth]{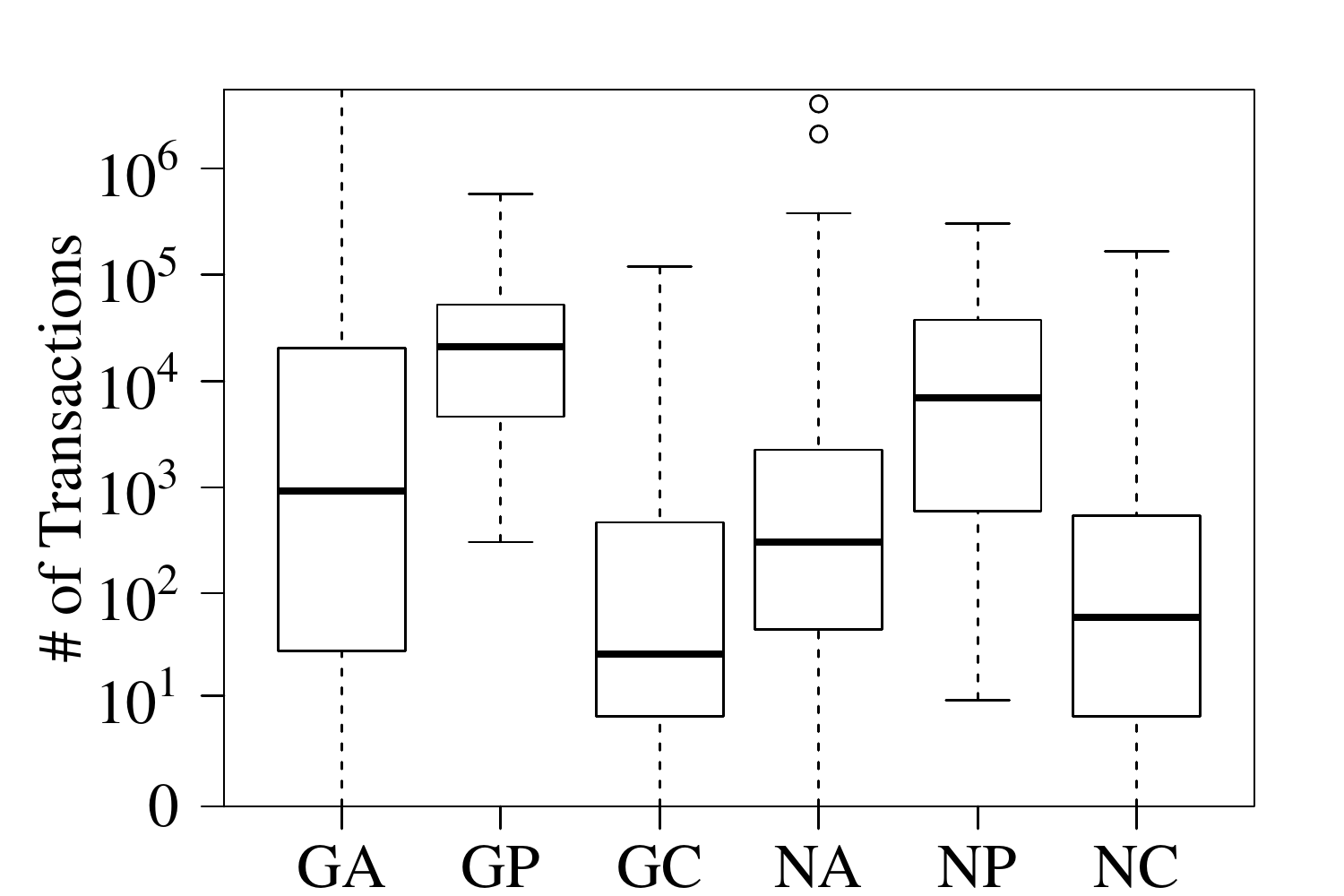}
    }
    \caption{Distribution of number of transactions among different open-source categories.}
    \label{fig:open_src_txn}
    \end{minipage}
\end{figure}

Figure \ref{fig:open_source} depicts the percentage of DApps in each open-source category. Only 15.7\% of DApps are fully open source, and 25.0\% of DApps are fully closed source. 54.9\% of DApps do not make them project-level open-source, meaning that they hide their business-related code. As for the contract-level open source, we can observe that there are 43.5\% of DApps whose smart contracts are all open source. This result implies that developers tend to share the code of smart contracts rather than the DApps.

Then we analyze the relationship between open source and popularity of DApps. Figure \ref{fig:open_src_txn} depicts the distribution of the number of transactions among different open-source categories. We can find that DApps whose smart contracts are all closed source have smaller number of transactions. DApps whose smart contracts are all open source have higher maximum transactions.
Therefore, we can conclude that the open source of smart contracts may improve the DApps' popularity.

\subsection{Usage Patterns of Smart Contracts}
\label{sec:patterns}

\begin{figure}[htbp]
    \centerline{
    \includegraphics[width=0.45\linewidth]{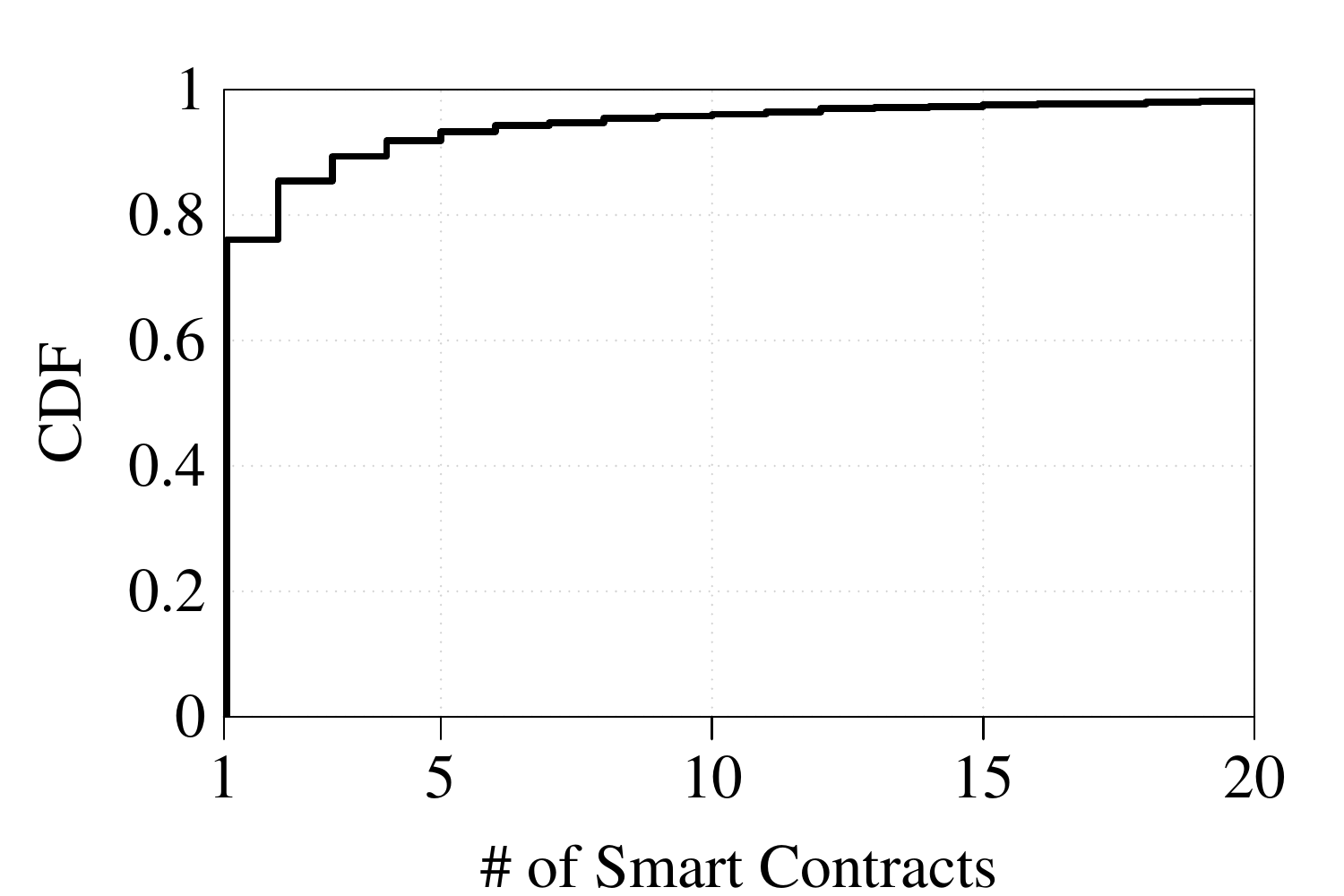}
    }
    \caption{Distribution of the number of smart contracts per DApp.}
    \label{fig:dapp_sc_cdf}
\end{figure}

Developers use smart contracts to keep data on the chain and do some operations. Sometimes the operations are too complex to be done by one smart contract, so that developers implement multiple smart contracts for a single DApp.
Figure \ref{fig:dapp_sc_cdf} shows the distribution of the number of smart contracts per DApp. We can find that over 75\% (757) of DApps are supported by one smart contract. We denote these DApps as single-contract DApps.

About 25\% of DApps are supported by multiple smart contracts (denoted as multi-contract DApps), and almost all of them have less than 10 smart contracts.
In the median case, a multi-contract DApp has 3 smart contracts. The maximum number of smart contracts of a DApp is 1881.

\begin{table}[htbp]
\newcommand{\tabincell}[2]{\begin{tabular}{@{}#1@{}}#2\end{tabular}}
\caption{Classification of usage patterns of smart contracts.}
\begin{center}
\begin{tabular}{lll}
\hline
& \textbf{\tabincell{l}{Deployed by\\user accounts}} & \textbf{\tabincell{l}{Deployed by\\smart contract\\accounts}} \\
\hline
\textbf{\textit{\tabincell{l}{There are\\internal transactions}}} & Leader-Member Pattern & N\/A \\
\hline
\textbf{\textit{\tabincell{l}{There are no\\internal transactions}}} & Equivalent Pattern & Factory Pattern \\
\hline
\end{tabular}
\label{tab:usage_patterns}
\end{center}
\end{table}

\begin{figure}[htbp]
    \centerline{
    \includegraphics[width=0.95\linewidth]{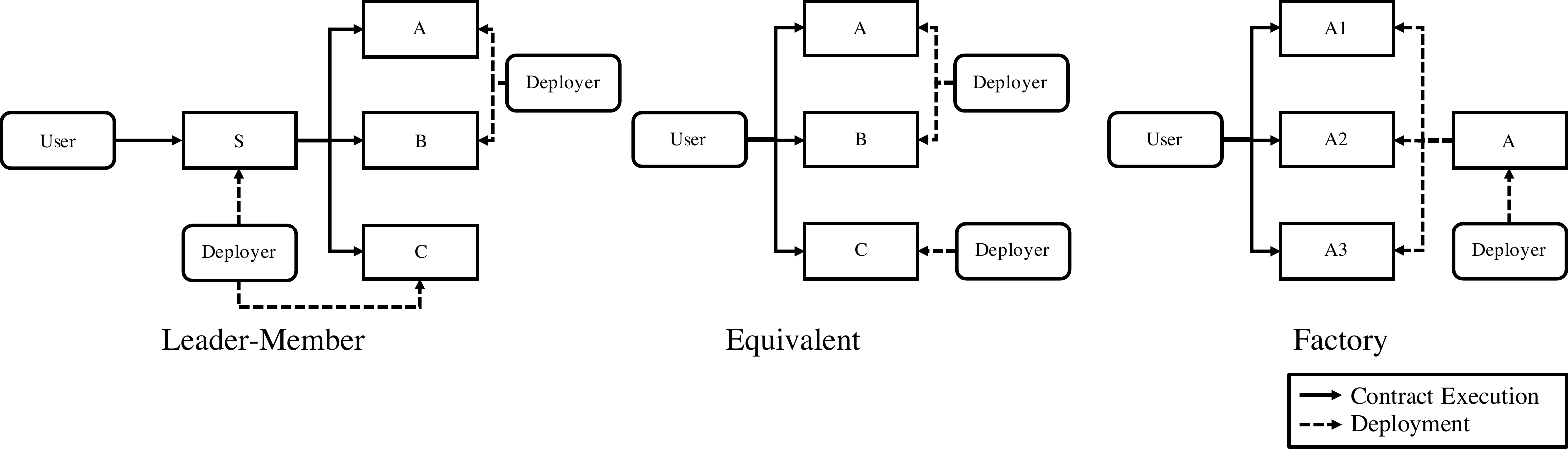}
    }
    \caption{Usage patterns of smart contracts.}
    \label{fig:usage_patterns}
\end{figure}

Considering features of smart contracts, we select two metrics to classify the usage patterns of smart contracts in multi-contract DApps: internal transactions among smart contracts and smart contract deployers. As shown in Table \ref{tab:usage_patterns}, we divide the smart contract usages into three patterns. Smart contract management in these patterns are illustrated in Figure \ref{fig:usage_patterns}. Internal transactions indicate that there is dependence among smart contracts. But internal transactions initiated by smart contracts generated by smart-contract accounts are used to return Ether to their deployers where they have no dependence in functionality. So, we do not need to create a category to describe this usage pattern.

\textbf{Leader-member pattern.}
In the leader-member pattern, a contract execution starts in a smart contract. Then the entry smart contract (the leader) initiates internal transactions to other smart contracts (members). Therefore, in the leader-member pattern, contract executions usually have more deeper call stacks. For a example, developers design smart contracts S, A, B, C, and set S as the entry contract. In a possible path, S is executed first, and then it sends internal transactions to A, B, C in order. So, S is the leader, A, B and C are members, and they form the leader-member pattern.

\textbf{Equivalent pattern.}
The equivalent pattern is the simplest pattern. Developers design smart contracts separately, make them handle contract executions, and combine their functions in clients and/or servers. By only transactions on the blockchain, we cannot find dependence among them. Like smart contract A, B, C deployed by different deployers, they don't send internal transactions to each other. But they support the DApp together.

\textbf{Factory pattern.}
In this pattern, a deployer deploys a smart contract and then make it deploy similar contracts to receive contract executions from user accounts. The process is like a factory producing products, so we call the pattern "factory pattern". The smart contract deployed first is the "factory contract", and smart contracts generated by the factory contract are called "child contracts". Figure \ref{fig:usage_patterns} shows a case: deployer deploy smart contract A first, and make it deploy A1, A2, A3. Then the three child contracts are able to be executed.
Developers using the factory pattern can keep child contracts similar.
This pattern is usually used in DApps from the category Gambling, whose developers use the factory contract to generate games with the same rules.

By checking internal transactions and deployers of smart contracts, we find leader-member pattern in 194 DApps and 199 smart contracts, equivalent pattern in 214 DApps and 1,539 smart contracts, and factory pattern in 28 DApps and 2,671 smart contracts.
The leader-member pattern is not widely used. Developers are more likely to design independent smart contracts in functionality.







\section{Cost of Smart Contracts in DApps}
\label{sec:sc}

The cost of smart contracts in DApps includes two parts: deployment cost and execution cost. Deployments and executions are done as transactions, which cost gas. Gas are paid with Ethers, and the amount of gas used is a measurement of the complexity of a contract execution. An account sends some gas in a contract execution, and then gets the gas left when the contract execution is confirmed. If the transaction has used all the gas sent from the initiator, the account receives an error information "out of gas" and lose all gas it sends.

To lower the costs of deployments and executions is important. In Ethereum blockchain, total gas of a block is limited. A complex smart contract may cost too much gas so that it cannot be deployed, i.e., the block will not contain the transaction. In addition, the higher the contract execution costs are, the lower the throughput of contract executions, and the longer users wait for confirmations of executions.

In this section, we investigate the gas that is actually used for deploying and executing smart contracts to answer RQ3, i.e., how much is the cost of DApps when running on the blockchain? If not specified, in this section, the term ``gas'' represents the amount of gas that has been actually used in the contract execution.

\subsection{Deployment Cost of Smart Contracts}
\label{sec:deploy_cost}

As mentioned above, gas used in deploying a smart contract reflects the complexity of the contract. We compare the deployment costs among different usage patterns of smart contracts to study whether developers can leverage different patterns to lower the cost of deploying smart contracts. Then we build a regression model to explore what factors influence the deployment cost.

\subsubsection{Deployment Cost of Different Usage Patterns of Smart Contracts}

\begin{figure}[htbp]
	\centerline{
    \includegraphics[width=0.95\linewidth]{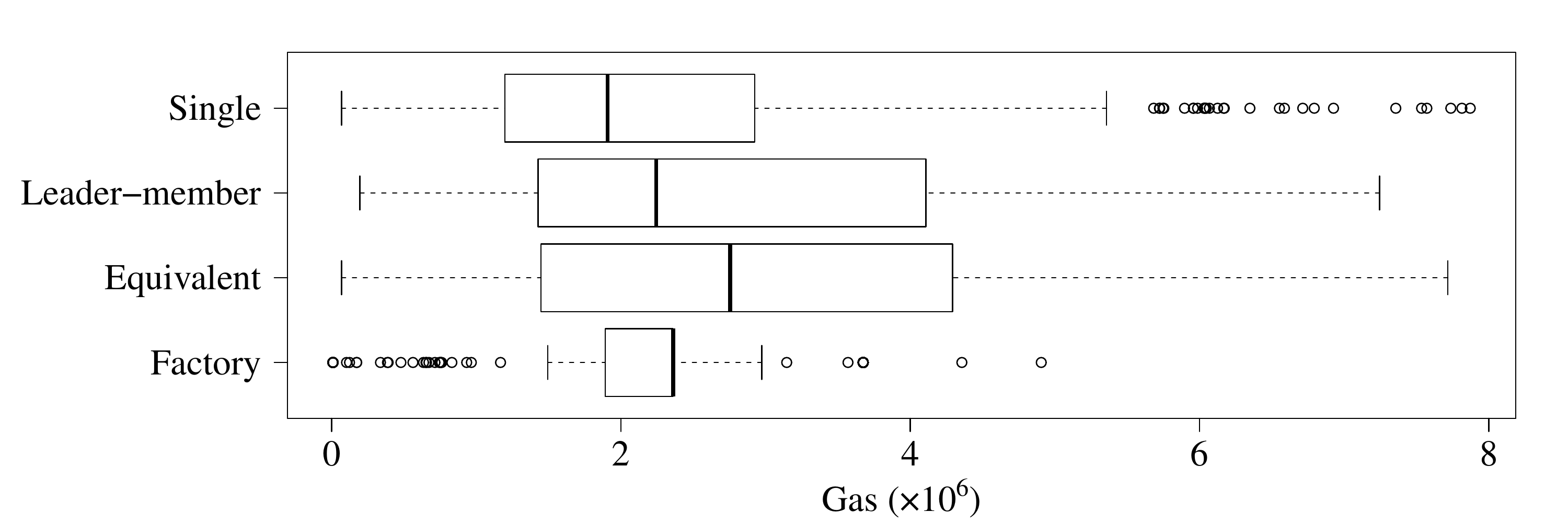}
    }
	\caption{Deployment cost difference among smart contract usage patterns.}
	\label{fig:deploy_diff}
\end{figure}

Figure \ref{fig:deploy_diff} shows the distribution of smart contract deployment costs among different usage patterns. In general, smart contracts of the single pattern cost smaller than others, meaning that the smart contracts of single-contract DApps are simpler than those of multi-contract DApps. The reason is that developers of single-contract DApps design limited functions in smart contracts, and may use back-end servers or relatively complex clients to implement the DApps.

Smart contracts of leader-member pattern and equivalent pattern have more deployment cost, meaning that contracts of these two patterns are more complex. The reason is that developers need smart contracts to do more operations. They split functions into multiple smart contracts and make them co-operate so that they could keep deployment cost of each smart contract at a low level.

Deployment costs of child contracts of the factory pattern are more concentrated around a value, because they are generated by some factory contracts. Child contracts generated by a certain factory contract have similar deployment costs.

\subsubsection{Influence Factors}

To explore what influences the deployment costs of smart contracts, we extract some features from smart contracts and their source code, and build a regression model to check relationships between features and the deployment cost.

From \textit{Etherscan}, we can get ABIs (Application Binary Interface) and source code of smart contracts. The ABI describes a smart contract, including functions (normal functions and constructors) and events. The number of functions (NoF for short) and lines of code (LoC for short) could reflect the complexity of a smart contract and may influence the deployment cost.

\begin{table*}[htbp]
\caption{Result of two linear regression analyses.}
\begin{center}
\begin{tabular}{lrrrrr}
\hline
\textbf{Feature} & \textbf{Estimate} & \textbf{Std. Error} & \textbf{$Pr(>|t|)$} & \textbf{Multiple R-squared} & \textbf{P-value} \\
\hline
\textbf{\textit{NoF}} & 53389 & 1965 & $<2e-16$ & 0.2956 & $<2.2e-16$ \\
\textbf{\textit{LoC}} & 1870 & 53.86 & $<2e-16$ & 0.4066 & $<2.2e-16$ \\
\hline
\end{tabular}
\label{tab:regression}
\end{center}
\end{table*}

First, each feature is separately used to build a linear regression model with deployment cost and do regression analysis by R \cite{rpro}. The result is shown in Table \ref{tab:regression}. At the level of t<0.01, the deployment cost is obviously related to these features.

\begin{table}[htbp]
\caption{Result of multiple regression analysis.\\(Multiple R-squared: 0.535, P-value: $<2e-16$)}
\begin{center}
\begin{tabular}{lrrr}
\hline
\textbf{Feature} & \textbf{Estimate} & \textbf{Std. Error} & \textbf{$Pr(>|t|)$} \\
\hline
\textbf{\textit{NoF}} & 37141.26 & 1685.74 & $<2e-16$  \\
\textbf{\textit{LoC}} & 1514.68 & 50.34 & $<2e-16$ \\
\hline
\end{tabular}
\label{tab:multiple_regression}
\end{center}
\end{table}

Then we use all features with the deployment cost to do multiple regression analysis (Table \ref{tab:multiple_regression}). $R^2$ is significantly higher than results of two linear regression analyses above. The correlation coefficient of NoF is higher than LoC's, so we can conclude that NoF and LoC all influence the deployment cost, and the deployment cost is more related to NoF.

\subsection{Execution Cost of Smart Contracts}
\label{sec:exec_cost}

In a contract execution, a user (the caller) sends some gas, input data and Ethers if they want, to call a function of the smart contract. Miners receive the transaction and add it into a block. When checking if the block is valid, miners will find the transaction and then execute it. They load the smart contract and its storage, and execute the function. They will load other smart contracts if the function calls other functions of other smart contracts, namely create internal transactions. They count gas for any instructions. If all the gas is used or another error occurs, miners will stop the execution and mark the contract execution as "isError", meaning some errors in the execution and the contract execution fails. Unless all gas is used, gas left will be changed into Ethers and returned to the caller after the block is added into the chain, namely the transaction is confirmed.

Therefore, to lower the costs of contract executions, callers should better send less gas in transactions. Meanwhile, developers should lower the complexity of contract executions. We check the gas sent and used in transactions, and explore if internal transactions and the usage pattern of smart contracts may influence the execution costs, to find a way to lower the cost.

\subsubsection{Gas Sent in Transactions}

In practice, end-users interact with clients, where some operations are translated to contract executions and sent to smart contracts. Contract executions (excluding the deployment) represent actual use of smart contracts. Users (sometimes the DApp maintainers) pay for them with Ethers. The gas left of a contract execution cannot be used until the transaction is confirmed. If users want to use their Ethers more flexibly, they should better send limited but enough gas.

\begin{figure}[htbp]
	\centerline{
	\subfloat[Gas left of a contract execution.]{
        \includegraphics[width=0.45\linewidth]{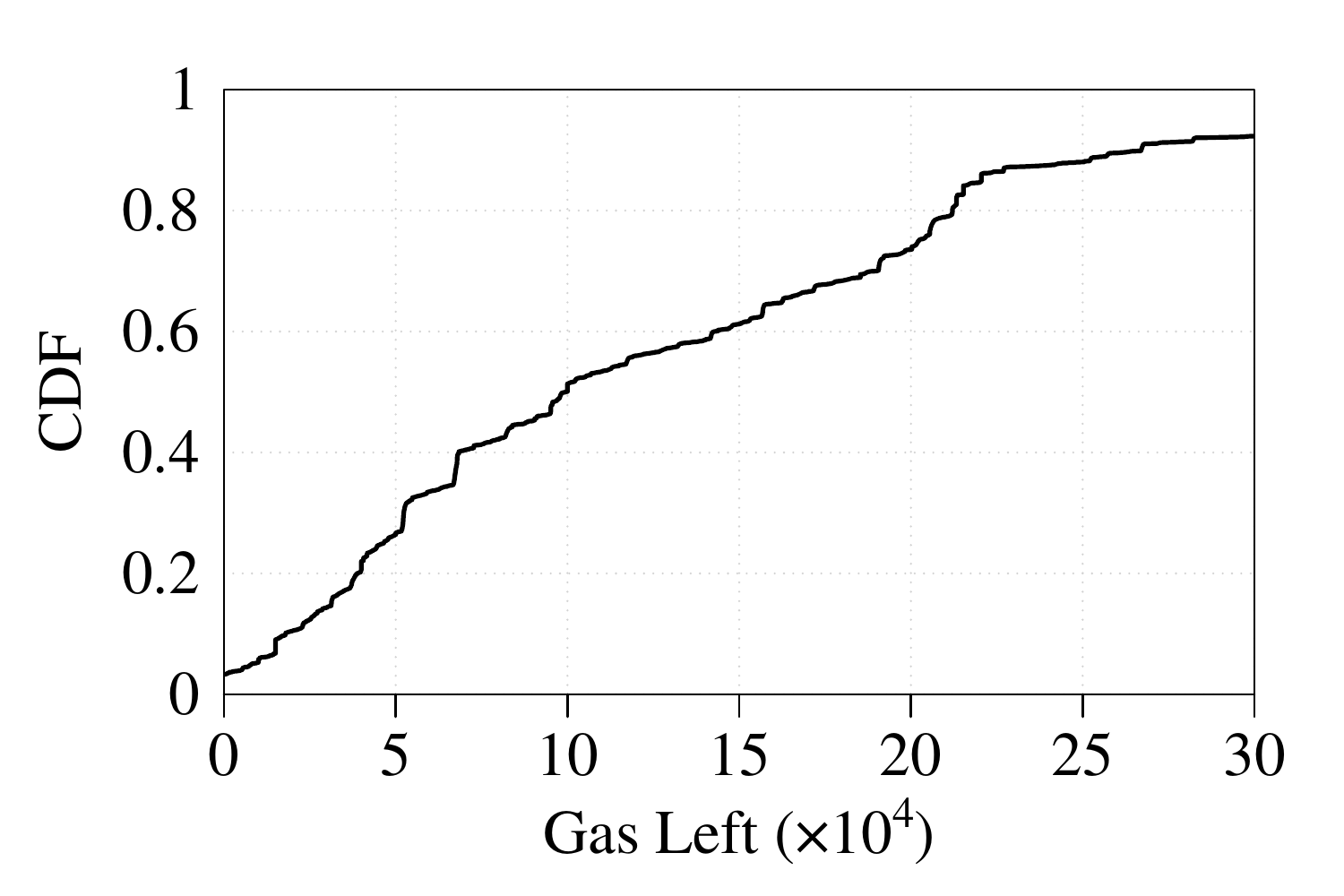}
        \label{fig:gas_left}
    }
    \subfloat[Gas used of a contract execution.]{
        \includegraphics[width=0.45\linewidth]{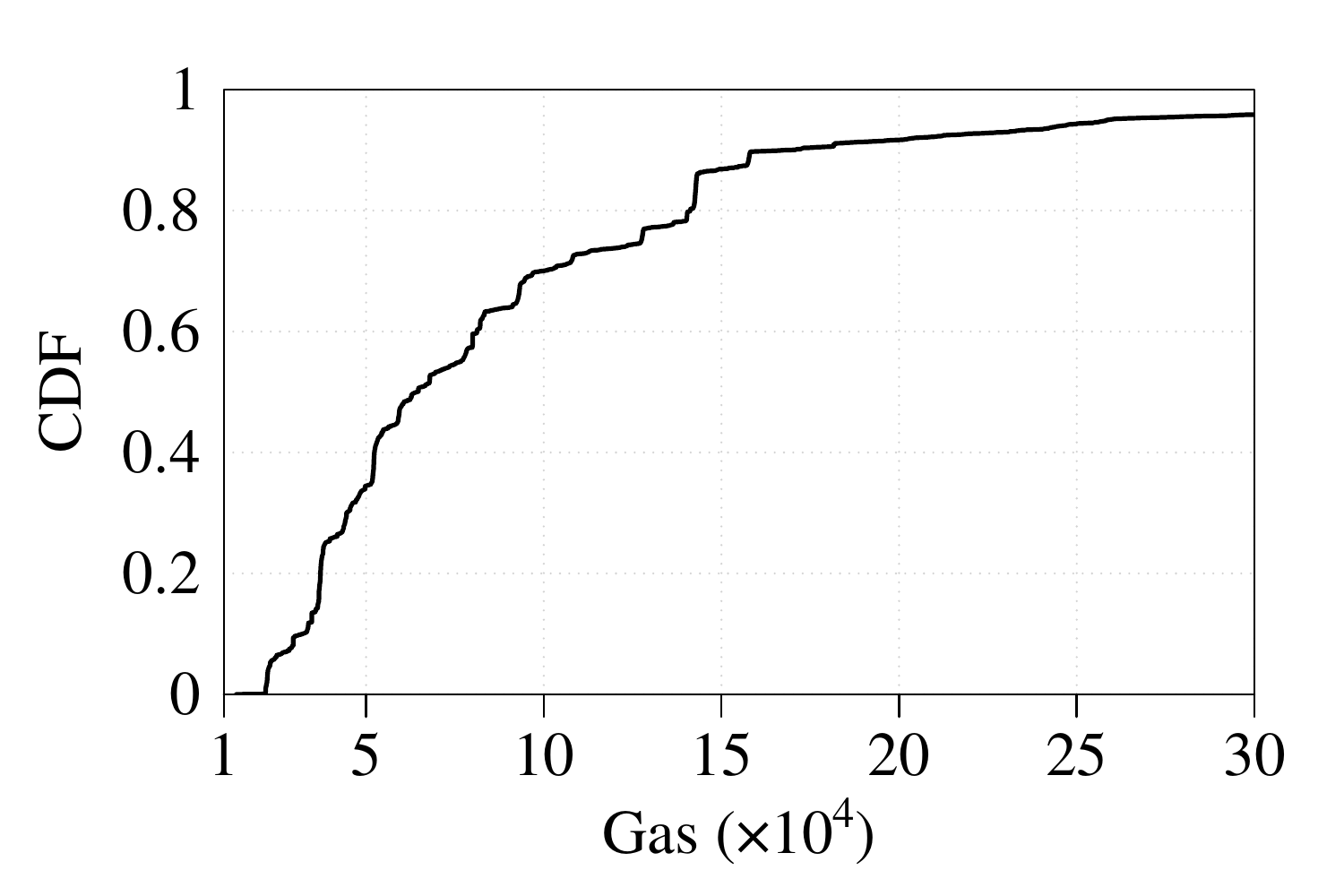}
        \label{fig:gas_used}
    }}
	\caption{Gas of contract executions.}
	\label{fig:gas}
\end{figure}

Figure \ref{fig:gas_left} shows the distribution of number of contract executions to gas left. We can see that 50\% of contract executions have 100,000 gas left and 75\% of contract executions have 200,000 gas left. About 5\% of contract executions use all the gas. Such results indicate that over 100,000 gas are locked in about 50\% of contract executions.

In Figure \ref{fig:gas_used}, we can find that 70\% of contract executions cost less than 100,000 gas, and 80\% of contract executions cost less than 141,213 gas. Only about 5\% of contract executions cost over 300,000 gas. So, if users want to request a contract execution, they can just send 141,213 gas to cover the cost in 80\% of cases.\\

In the median case, users send 199,366 gas in contract executions, but only 99,366 gas is used and 100,000 gas left is locked in contract executions until transactions are confirmed.

\subsubsection{Contract Executions with Internal Transactions}

Internal transactions also cost gas. Because smart contracts cannot actively initiate internal transactions, they are triggered by external transactions. Internal transactions are included in external transactions that trigger them and the costs are included in costs of contract executions as well. Thus, for similar contract executions, those that have internal transactions cost more gas than others.

Internal transactions are initiated for many reasons, but just a few of them are in design of DApps. Suppose there are two DApps, A and B, each of which is a multi-contract DApp. A smart contract of A is triggered by a contract execution to initiate an internal transaction to B, and then there are more internal transactions triggered among smart contracts of B. Internal transactions among smart contracts of B are not taken into account for developers of A, and they do not have to think about them. Therefore, we just consider internal transactions in DApp designs, which means internal transactions among smart contracts of DApps.

\begin{figure}[htbp]
    \centerline{
    \includegraphics[width=0.45\linewidth]{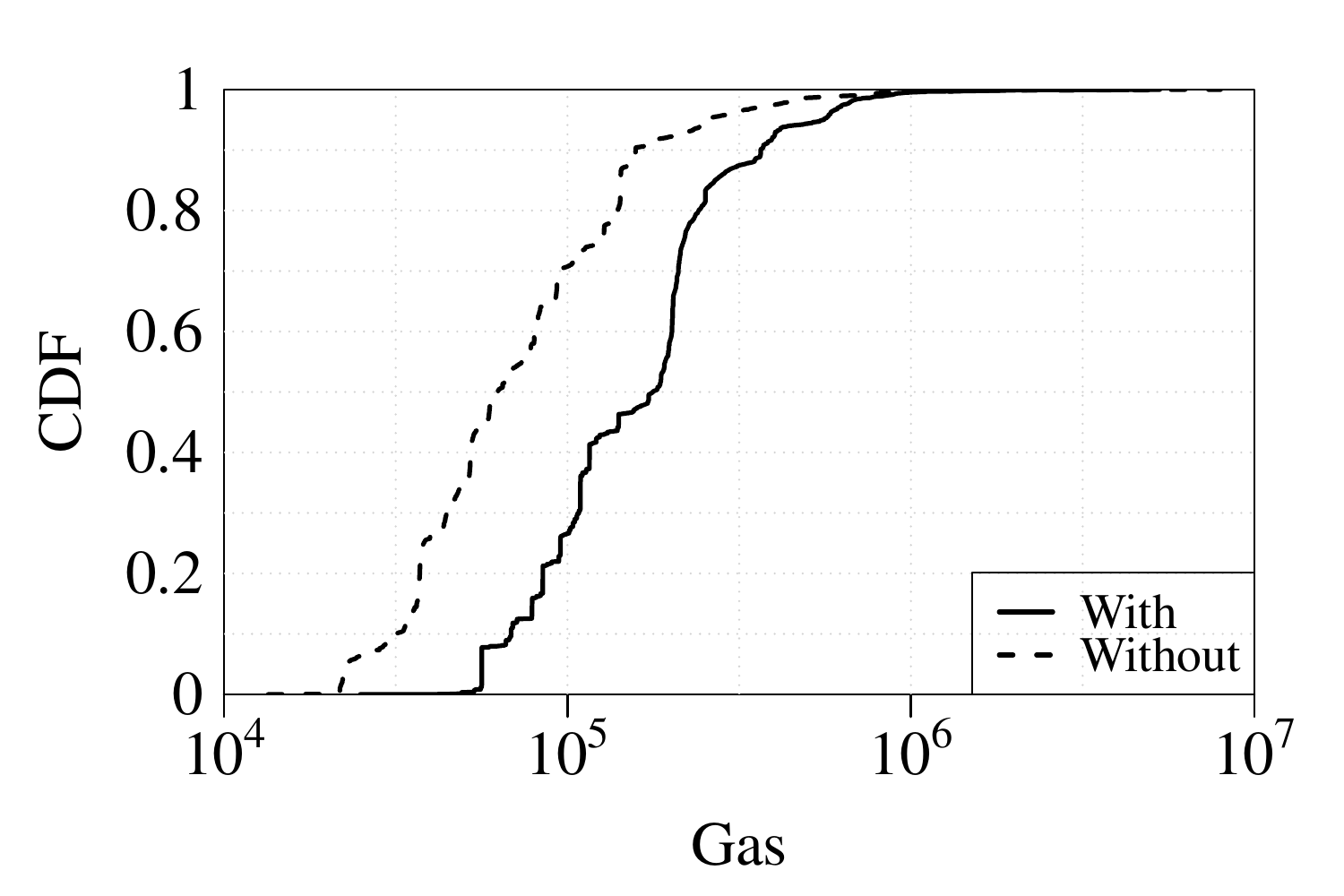}
    }
    \caption{Execution cost difference between contract executions with and without internal transactions in DApps.}
    \label{fig:gas_it_difference}
\end{figure}

Figure \ref{fig:gas_it_difference} shows difference between the two types of contract executions. We can find that contract executions with internal transactions cost more gas than those without internal transactions.

\subsubsection{Difference among Smart Contract Usage Patterns}

To compare execution costs of smart contracts among usage patterns, we have to use a metric to represent the average level of a smart contract's execution costs. We select an example from our dataset to show the distribution of a contract's contract executions and their costs.

\begin{figure}[htbp]
    \centerline{
    \subfloat[Gas of a contract execution.]{
        \includegraphics[width=0.45\linewidth]{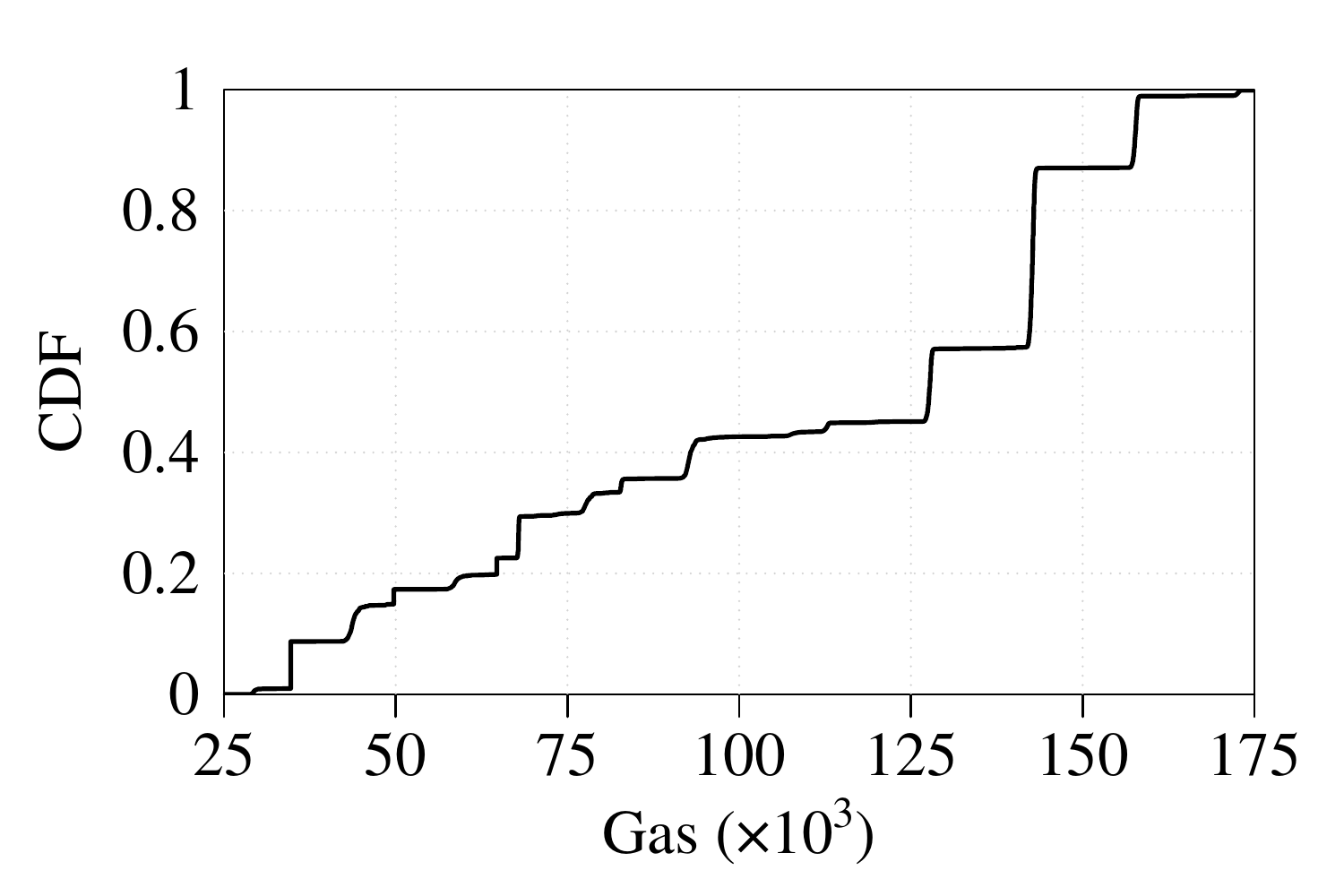}
        \label{fig:largest_gas_cdf}
    }
    \subfloat[Distribution of numbers of invocations to functions]{
        \includegraphics[width=0.45\linewidth]{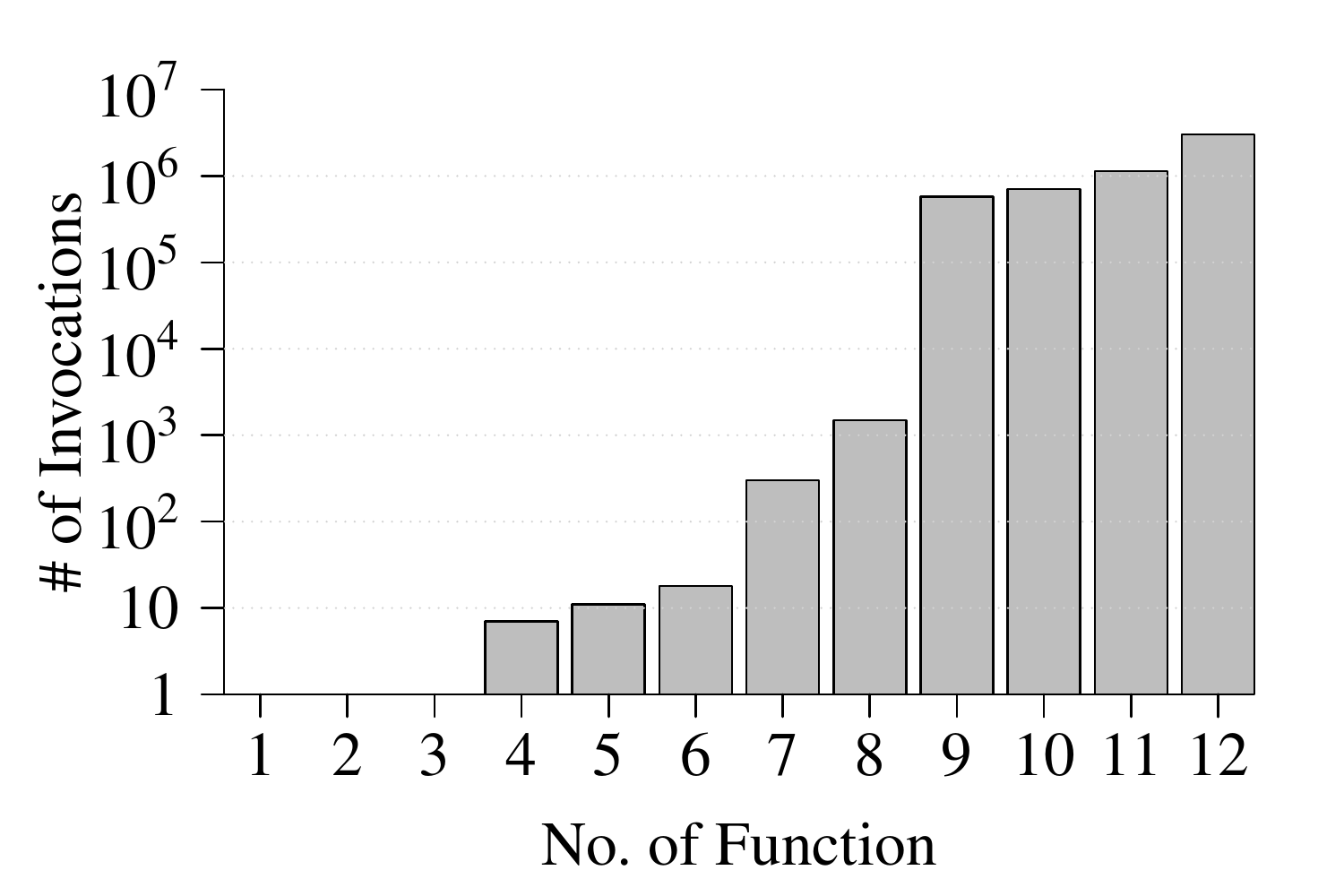}
        \label{fig:largest_invocation}
    }
    }
    \centerline{
    \subfloat[Distribution of gas used of each function]{
        \includegraphics[width=0.45\linewidth]{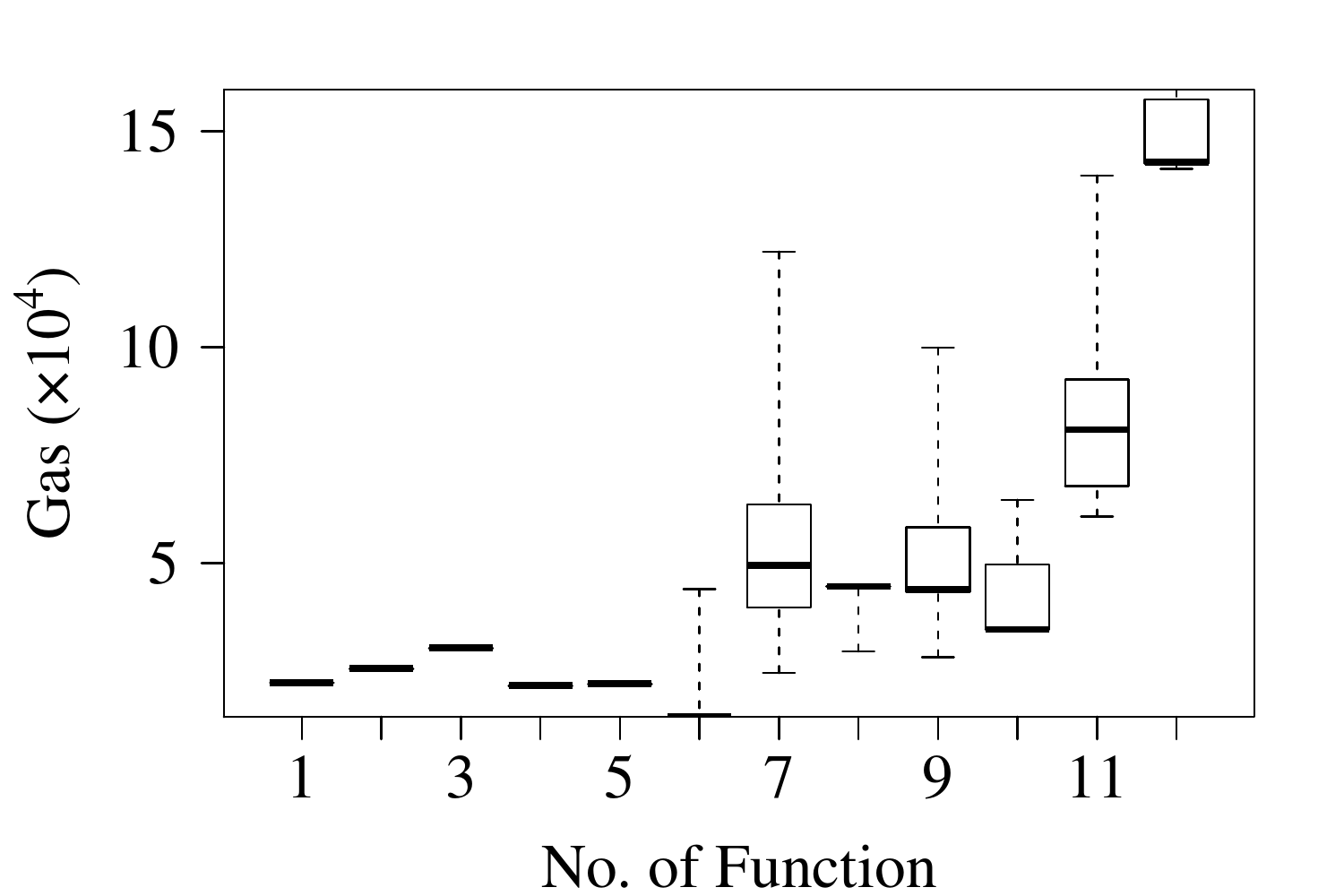}
        \label{fig:largest_method_gas}
    }
    }
    \caption{Contract 0x2a0c0DBEcC7E4D658f48E01e3fA353F-44050c208 of IDEX.}
    \label{fig:largest_contract}
\end{figure}

IDEX \cite{idex} is a DApp categorized as Exchanges and has the most transactions in 2018. The contract selected is one of the two smart contracts it has. As shown in Figure \ref{fig:largest_contract}, distributions of the contract's execution costs and numbers of invocations of functions are uneven. Neither arithmetical mean nor median can represent the average level.

For each function, gas used of its invocations is concentrated. So we take the number of invocations to functions into account, and define a new metric $agas$ (average gas) to represent the average level of a smart contract's execution cost:

\begin{equation}
agas=\frac{\sum mgas_i\cdot count_i}{\sum count_i}
\end{equation}

$mgas_i$ represents gas median of function i, $count_i$ represents the invocation count of function i.

\begin{figure}[htbp]
    \centerline{
    \includegraphics[width=0.95\linewidth]{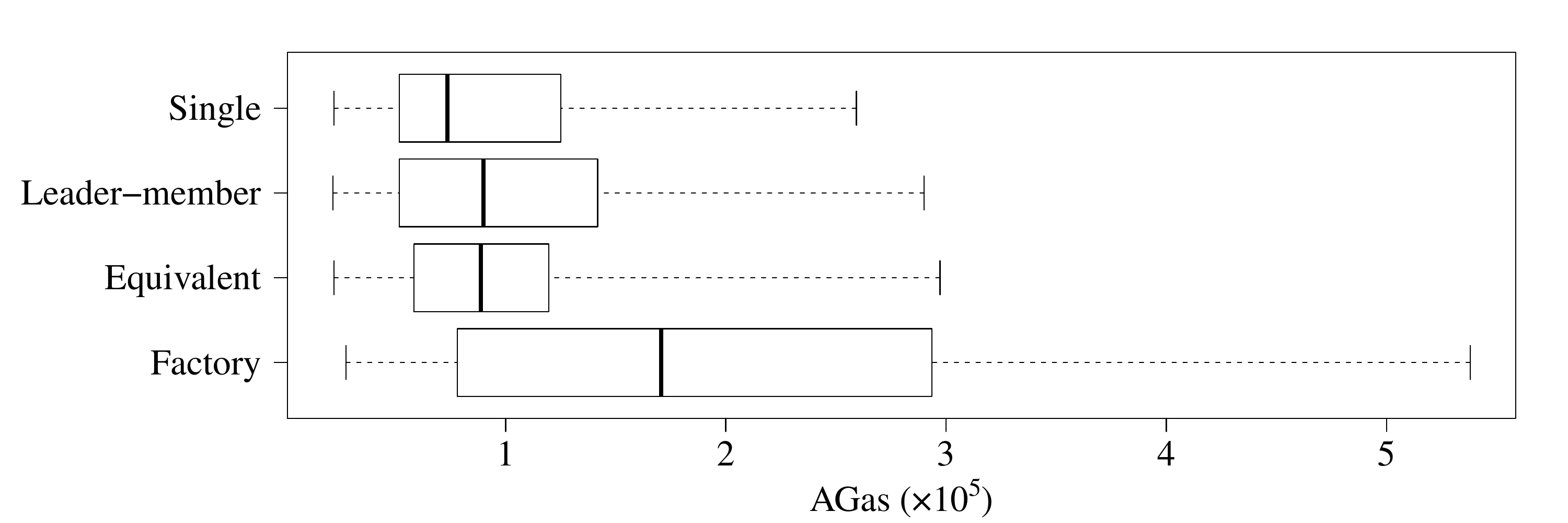}
    }
    \caption{Execution cost difference among smart contract usage patterns.}
    \label{fig:gas_difference}
\end{figure}

Then we compare the average gas of smart contracts of four patterns.
According to Figure \ref{fig:gas_difference}, in general, smart contracts of leader-member pattern and factory pattern cost more gas in contract executions. Smart contracts of equivalent pattern have a similar distribution of $agas$ with those of single pattern.
\section{Findings and Implications}
\label{sec:implications}

Based on the results in preceding sections, we summarize the findings of our study and draw implications to the stakeholders in the ecosystem of blockchain-based DApps, including end-users, DApp developers, and blockchain vendors. Table \ref{tab:findings_implications} shows all the findings and implications.

\begin{table*}[htbp]
\caption{Summary of findings and implications.}
\begin{center}
\begin{tabular}{p{0.35\linewidth}p{0.35\linewidth}p{0.2\linewidth}}
\hline
\textbf{Findings} & \textbf{Implications} & \textbf{Stakeholders} \\
\hline
\textbf{\textit{F1.}}
On the blockchain such as Ethereum, DApps with financial features are more popular than others, such as DApps from the Exchanges, Finance, and Gambling categories. In contrast, other categories have relatively less users and transactions. &
The development of blockchain-based DApp market is still at an early stage, mainly aiming at meeting the financial requirements. DApps are not popular to satisfy more general application contexts. &
End-users\\
\hline

\textbf{\textit{F2.}}
Other categories can become popular, too, like Games. All DApps from these categories have financial features, such as in-app purchase services or collectible cards bought with Ethers. &
Adding financial features, like in-app purchase functions, may make DApps more popular. &
DApp developers \\
\hline

\textbf{\textit{F3.}}
As more DApps are developed, the number of DApps from the high-risk category increases significantly. &
Users have to be more careful to avoid Ponzi schemes or other similar investment frauds, even in DApps that seem not to be related to investment and gambling. &
End-users\\
\hline

\textbf{\textit{F4.}}
Only 15.7\% of DApps are fully open source where the source code of the project and smart contracts is published. 52.3\% of DApps' smart contracts are closed source. &
DApps does not obey the blockchain's principles on opening and auditability. Blockchain vendors should better provide mechanisms to ensure the open source of DApps. &
Blockchain vendors \\
\hline

\textbf{\textit{F5.}}
DApps whose smart contracts are open source have more transactions than those whose smart contracts are all closed source. &
It is better for developers to make the smart contracts open source, which could improve the popularity of DApps. &
DApp developers\\
\hline

\textbf{\textit{F6.}}
About 75\% of DApps have only one smart contract. For multi-contract DApps, most of the contracts are independent where only 199 (3.9\%) smart contracts have internal transactions. &
Developers should be careful in development. Because smart contracts can't be modified after deployments, if some smart contracts break down, developers of multi-contract DApps just need to replace the smart contract in which errors occur, but most developers have to redeploy all smart contracts of their DApps, namely the only one smart contract of single-contract DApps. &
DApp developers \\
\hline

\textbf{\textit{F7.}}
In general, deploying a smart contract costs millions of gas. Both NoF (number of functions) and LoC (line of code) influence the deployment cost of smart contracts, and the deployment cost is more related to NoF. &
Reducing NoF and LoC may lead to lower deployment cost of smart contracts. Reducing the number of functions could help more. &
DApp developers\\ 
\hline

\textbf{\textit{F8.}}
50\% of contract executions have over 100,000 gas left. 80\% of contract executions cost less than 141,213 gas. &
To prevent more gas from being locked in contract executions, users can just send 141,213 gas. In 80\% of the cases, such an amount of gas can cover the cost. &
End-users and DApp developers\\
\hline

\textbf{\textit{F9.}}
Contract executions with internal transactions in DApps generally cost more gas than those without internal transactions. &
Splitting all functions into independent parts and using the equivalent pattern to organize smart contracts may reduce the execution cost. If functions are too complex to be split, it is better to keep just key functions in smart contracts and use a service to combine them on a server. &
DApp developers\\

\hline
\textbf{\textit{F10.}}
Some developers use factory pattern in DApps, i.e., deploying a factory contract which is used to deploy new child smart contracts. Child contracts have more concentrated deployment costs, and generally have higher execution costs. &
Because the factory contract must have the code of child contracts, complex child contracts lead to high deployment cost. If so, equivalent pattern may help to avoid extra deployment cost of the factory contract. &
DApp developers\\
\hline
\end{tabular}
\label{tab:findings_implications}
\end{center}
\end{table*}
\section{Related Work}
\label{sec:related_work}

To the best of our knowledge, our work is the first comprehensive study to understand the blockchain-based DApps. In this section, we first survey the related work of blockchain systems, and then introduce literature on P2P applications, which are the traditional type of DApps.

\subsection{Blockchain Systems}
The blockchain has become a hot research field. Researchers mainly focus on three research directions: the underlying mechanism, application and data mining.

\textbf{Underlying mechanism.}
Underlying Mechanisms include consensus mechanisms and smart contract mechanisms. Many consensus mechanisms are proposed, such as PoW (Proof of Work) \cite{nakamoto2008bitcoin}, PoS (Proof of Stake) \cite{king2017ppcoin}, DPoS (Delegated Proof of Stake) \cite{larimer2014delegated}, and Algorand \cite{gilad2017algorand} recently presented. Some classical consistency algorithms like PBFT \cite{barger2017scalable} (practical byzantine fault tolerance) are used as well.
Furthermore, a few researchers make performance monitoring \cite{gervais2016security, zheng2018detailed} and try to improve existing mechanisms \cite{zhang2017rem, chen2017adaptive}. Smart contract mechanisms attract experts of software engineering \cite{liao2017toward, porru2017blockchain} and security \cite{luu2016making, kosba2016hawk, coblenz2017obsidian}.

\textbf{Blockchain application.}
For its decentralization, persistency, anonymity and auditability, blockchain can be used in anonymous trading, persistence services and cross-organizational transactions. So blockchain technology has been widely used in finance service \cite{klems2017trustless}, IoT (Internet of Things) \cite{shae2017design, ruta2017supply, bahri2018trust}, information security \cite{liu2017blockchain, nguyen2018trusternity}, edge computing \cite{xu2019blockchain} and software engineering \cite{liao2017toward}.

\textbf{Data mining.}
Thanks to the public accessibility and auditability of blockchain, researchers can do analysis based on transaction data for usage characteristics \cite{ron2013quantitative, meiklejohn2013fistful, chen2018understanding} and so on.
The other accessible data are smart contracts, for example, their bytecode written in the blocks. There are some work analyzing these code to give advises for smart contract developers and blockchain users \cite{chen2018detecting, chen2017under, chang2018scompile}. Researchers also try to decompile them into source code \cite{grech2019gigahorse} so that more approaches to source code analysis can be used.

\subsection{Peer-to-peer Applications}

Traditional DApps refer to applications on the P2P network \cite{whatisadapp}, on which there are lots of research efforts, including security, performance and application.

\textbf{Security.}
Security of P2P network and applications includes two parts: security of P2P network and possible harmful behaviours of users. For security of P2P network, researchers detect attacks \cite{locasto2005towards}, use other technologies like trusted computing \cite{zhang2005enhancing} and design new protocols \cite{cooper2009flip}. Some harmful behaviours in P2P applications are found as well \cite{liang2005pollution, lian2007empirical}.

\textbf{Performance.}
P2P applications are likely to become the burden of local network, especially P2P file sharing systems. Researchers monitor the performance \cite{saroiu2001measurement, pouwelse2005bittorrent, abeni2019demonstrating}, and try to improve these applications by optimizing application layer \cite{loo2004enhancing, nurminen2013p2p, amad2015self} and network layer \cite{liu2010cross}.

\textbf{Application.}
P2P technology is widely used in many fields, such as instant messaging \cite{bitmessage}, file sharing system \cite{napster}, development \cite{briola2019platform}, security \cite{friedman2015generic} and so on.

Cryptocurrencies, for example Bitcoin, are applications based on P2P networks as well. From 2009, more and more cryptocurrencies are issued, which leads to the growth of blockchain technology, especially public blockchain technology. According to CoinMarketCap \cite{coinmarketcap}, nowadays there are over 2,000 cryptocurrencies in the world.

Few researches are about blockchain-based DApps, but these DApps have great influence on the blockchains on which they run. Some work has been performed to help developers develop Blockchain-based DApps \cite{dong2019proofware}.
\section{Conclusion and Future Work}
\label{sec:conclusion}

The scale of DApp market is rapidly growing and has reached billions of dollars. In this study, we conducted a systematic descriptive analysis of 995 DApps over Ethereum.
We analyzed the popularity of DApps, the development practices of DApps, and the cost of running DApps. Our findings provided valuable implications for different stakeholders in the ecosystem of blockchain-based DApps, including end-users, DApp developers and blockchain vendors.

This paper mainly focused on the descriptive analysis of DApps' transaction data. In our future work, we plan to study how to develop a high-quality DApp by deeply investigating the source code of DApps. Some questions needed to be further explored, such as how to build a DApp project, how to keep synchronization on and off the chain, and how to improve the throughput of DApps. The answer of these questions can directly improve the development of DApps and benefit millions of users.

\bibliographystyle{./bibliography/IEEEtran}
\bibliography{main.bib}


\end{document}